\documentclass[sigconf]{acmart}
\AtBeginDocument{%
  \providecommand\BibTeX{{%
    \normalfont B\kern-0.5em{\scshape i\kern-0.25em b}\kern-0.8em\TeX}}}

\copyrightyear{2023}
\acmYear{2023}
\setcopyright{acmlicensed}\acmConference[CHI '23]{Proceedings of the 2023 CHI Conference on Human Factors in Computing Systems}{April 23--28, 2023}{Hamburg, Germany}
\acmBooktitle{Proceedings of the 2023 CHI Conference on Human Factors in Computing Systems (CHI '23), April 23--28, 2023, Hamburg, Germany}
\acmPrice{15.00}
\acmDOI{10.1145/3544548.3580736}
\acmISBN{978-1-4503-9421-5/23/04}

\graphicspath{{figures/}{pictures/}{images/}{./}} % where to search for the images
\newcommand{\FingerMapper}{FingerMapper}
\newcommand{\Attach}{\textit{Attach}}
\newcommand{\Direct}{\textit{Direct}}
\newcommand{\raycastingCond}{\textit{ray-casting}}
\newcommand{\physicalCond}{\textit{hand tracking}}
\newcommand{\FingerSaber}{FingerSaber}
\newcommand{\FingerGrabber}{FingerGrabber}
\newcommand{\FingerClimber}{FingerClimber}

\usepackage{color}
\def\hl#1{#1}

\begin{document}

% \title{FingerMapper: Enabling Virtual Arms Movements in Confined Spaces by Re-Mapping Finger Motions in Virtual Reality}
\title{FingerMapper: Mapping Finger Motions onto Virtual Arms\\to Enable Safe Virtual Reality Interaction in Confined Spaces}

% \author{Anon.}
% \email{anon@anon.com}
% \affiliation{%
%   \institution{Anon.}
%   \city{Anon.}
%   \country{Anon.}
% }

\author{Wen-Jie Tseng}
\affiliation{%
  \institution{LTCI, INFRES, Telecom Paris, IP Paris}
  \city{Palaiseau}
  \state{}
  \country{France}
}
\email{wen-jie.tseng@telecom-paris.fr}

\author{Samuel Huron}
\affiliation{%
  \institution{CNRS i3 (UMR 9217), SES, Telecom Paris, IP Paris}
  \city{Palaiseau}
  \state{}
  \country{France}
}
\email{samuel.huron@telecom-paris.fr}

\author{Eric Lecolinet}
\affiliation{%
  \institution{LTCI, INFRES, Telecom Paris, IP Paris}
  \city{Palaiseau}
  \state{}
  \country{France}
}
\email{eric.lecolinet@telecom-paris.fr}

\author{Jan Gugenheimer}
\affiliation{
  \institution{TU Darmstadt}
  \city{Darmstadt}
  \country{Germany}\\
}
\affiliation{%
  \institution{LTCI, INFRES, Telecom Paris, IP Paris}
  \city{Palaiseau}
  \state{}
  \country{France}
}
\email{jan.gugenheimer@tu-darmstadt.de}

\renewcommand{\shortauthors}{Tseng, et al.}

\begin{abstract}
Whole-body movements enhance the presence and enjoyment of Virtual Reality (VR) experiences. However, using large gestures is often uncomfortable and impossible in confined spaces (e.g., public transport). We introduce FingerMapper, mapping small-scale finger motions onto virtual arms and hands to enable whole-body virtual movements in VR. In a first target selection study (n=13) comparing FingerMapper to hand tracking and ray-casting, we found that FingerMapper can significantly reduce physical motions and fatigue while having a similar degree of precision. In a consecutive study (n=13), we compared FingerMapper to hand tracking inside a confined space (the front passenger seat of a car). The results showed participants had significantly higher perceived safety and fewer collisions with FingerMapper while preserving a similar degree of presence and enjoyment as hand tracking. Finally, we present three example applications demonstrating how FingerMapper could be applied for locomotion and interaction for VR in confined spaces.
\end{abstract}

\ccsdesc[500]{Human-centered computing~Virtual reality}

\keywords{FingerMapper; Confined Spaces; Body Re-Association in VR}

\begin{teaserfigure}
  \includegraphics[width=\textwidth]{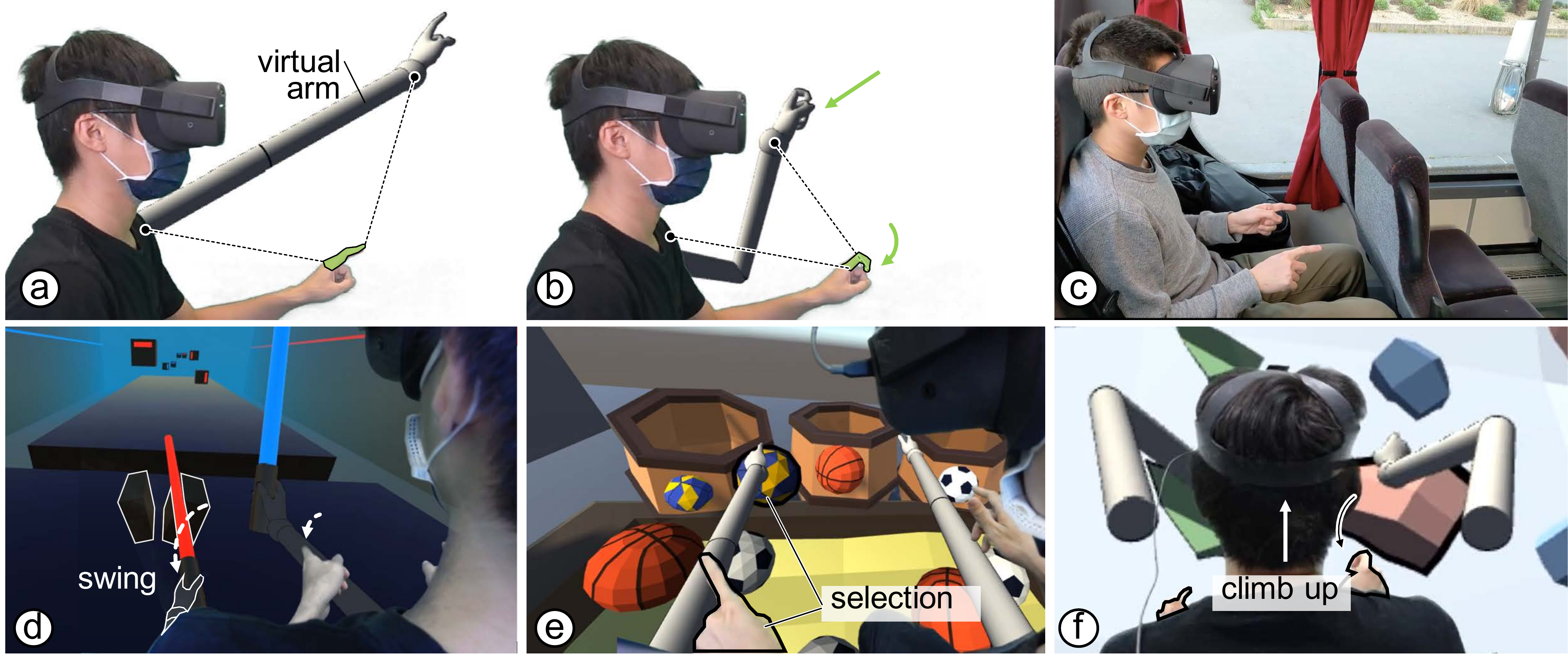}
  \caption{(a) We introduce \FingerMapper\ that leverages (b) small and less physically demanding finger motions to control a virtual arm. (c) We envision that the user can use FingerMapper for VR in confined spaces (e.g., the passenger seat of a bus) while allowing for virtual arm interactions with fewer collisions. 
  FingerMapper is available for a variety of interaction styles: (d) swinging motions in slicing-style games, (e) object interaction, and (f) locomotion in climbing-based games.}
  \Description{Figures 1a and 1b show FingerMapper, an interaction that uses small finger motions to control the whole-body scale virtual arm in VR. Figure 1c depicts that FingerMapper enables VR interaction when the user locates in confined spaces like commuting in a bus. FingerMapper can also be integrated with different VR interactions that are normally done by whole-body interaction, including arm-swinging motion (1d), object interaction (1e), and locomotion (1f).}
  \label{fig:teaser}
\end{teaserfigure}

\maketitle

\section{Introduction}
% whole-body motions enhances VR experiences
Current Virtual Reality (VR) experiences often require whole-body movement to increase presence \cite{slaterInfluenceBodyMovement1998, usohWalkingWalkinginplaceFlying1999}. Whole-body movement is a significant factor in engagement and sensory immersion \cite{isbisterGuidelinesDesignMovementBased2015}, suggested by research on physicality in games. Although having whole-body movement in VR enables a higher presence and enjoyment, these techniques also require a larger physical space \cite{marweckiVirtualSpaceOverloadingPhysical2018b}.

% mobile VR, VR in transports show a use case where whole-body motion is unavailable.
As VR Head-Mounted Displays (HMDs) are getting more mobile (e.g., Oculus Quest), users can immerse themselves wherever they wish. One prominent example would be using VR in transportation (e.g., planes \cite{williamsonPlaneVRSocialAcceptability2019}, cars \cite{hockCarVREnablingInCar2017, mcgillAmPassengerHow2017}, and others \cite{mcgillVirtualRealityPassenger2019, mcgillChallengesPassengerUse2020}). However, these scenarios often happen inside confined spaces \cite{gugenheimerChallengesUsingHeadMounted2019, liRearSeatProductivityVirtual2021}, where whole-body movement is usually undesirable since VR users do not have enough space for interaction and could even accidentally collide with bystanders or surroundings.

% In this paper, we present \FingerMapper---mapping small scale finger motions (Figure \ref{fig:teaser}ab) onto large scale virtual arm motions that enables embodied VR arm and hand interaction in public and confined spaces.
In this paper, we present \FingerMapper---a concept that maps small-scale finger motions onto large-scale virtual arm motions (Figure \ref{fig:teaser}ab). \FingerMapper\ enables VR experiences in confined spaces (e.g., the passenger seat in a bus, Figure \ref{fig:teaser}c) with fewer physical motions and collisions while preserving presence and enjoyment. Although confined spaces are unavailable for whole-body movements, we can benefit from the beyond-real interaction in VR \cite{abtahiBeingRealSensorimotor2022} by creating mapping metaphors between the VR user's finger and the virtual arm. Additionally, users can learn different relationships between tracked and rendered motion in VR to control an avatar \cite{wonHomuncularFlexibilityVirtual2015}. We implement two design variations (\emph{\Attach}\ and \emph{\Direct}) and present three applications (Figure \ref{fig:teaser}d-f) that support VR interaction, including large swinging motions (\FingerSaber), object interaction in space (\FingerGrabber), and locomotion (\FingerClimber).

% summary of study 1 and 2
To understand the usability of \FingerMapper, we first conducted a within-subject user study (n=13) including both design variations (\Attach\ and \Direct) with two baselines (\physicalCond\ and \raycastingCond) in a target selection task. While results show that \FingerMapper\ had a higher task completion time and lower Virtual Body Ownership (VBO) compared to \physicalCond, both mappings reduce the physical path length of participant's hands and fatigue. Next, we conducted a user experience study in a confined space (n=13) where participants played \FingerGrabber\ with \FingerMapper\ (\Attach) and \physicalCond\ inside the front passenger seat of a car. Results show that \FingerMapper\ had fewer collisions and a similar degree of presence and enjoyment to hand tracking. Overall, participants reported a significantly higher perceived safety using \FingerMapper\ in a confined space compared to hand tracking. Most participants (10 out of 13 in Study 2) reported they prefer using \FingerMapper\ in confined spaces to avoid colliding and keep feeling safe during the interaction. We see \FingerMapper\ as an alternative for scenarios where the user wants whole-body motion without having the necessary physical space for interaction. Finally, we discuss the potential scenarios of using \FingerMapper\ for future VR experiences (e.g., long-duration usage, context adaptive input).

% [Contributions]
Our work has three contributions. 1) The concept and implementation of \FingerMapper\ leveraging small finger motions onto virtual arms and hands to enable VR interaction in confined spaces. 2) Insights from two studies show that \FingerMapper\ has less physical movement, less fatigue, and fewer collisions compared to hand tracking. Participants also reported a similar degree of presence and enjoyment, a higher perceived safety, and preferred to use \FingerMapper\ in confined spaces (10 out of 13 in study 2). 3) The implementation of three applications (\FingerGrabber, \FingerSaber, and \FingerClimber) that demonstrate how to integrate \FingerMapper\ for different interaction scenarios.

%%%%%%%%%%%%%%%%%%%%%%%%%%%%%%%%%%%%%%%%%%%%%%%%%%%%%%%%%%%
% \section{Envisioned Scenario}
% VR users can switch to finger mapping techniques when the environment is constrained and does not allow large motions. \textit{``A user started playing BeatSaber at their home but had to meet a friend. During the commute on the bus, the user sat in a confined public space and still wanted to continue the game. On the system level, the user changed the tracking boundaries to be stationary and selected to have `confined' input. The VR HMD was now applying the finger mapping technique to map the user's hand motions to the virtual arms in the game.''} Here, our approach is not yet another interaction technique that has to be implemented to every application, but an abstract mapping that can be applied to any application working with hand tracking.

%%%%%%%%%%%%%%%%%%%%%%%%%%%%%%%%%%%%%%%%%%%%%%%%%%%%%%%%%%%
\section{Related Work}
Our work is related to confined spaces in VR and improving ergonomics for VR interaction. For the second subsection, we discuss prior research based on the framework of Abtahi et al. \cite{abtahiBeingRealSensorimotor2022}.

\subsection{Confined Spaces as a New VR Context}
The prevalence of VR products (e.g., HTC Vive, Oculus Rift) pushes the VR experience from a well-controlled environment (e.g., laboratory, demo booth) to the user's private space. Plenty of new contexts emerge while people use VR technologies in their homes. One prominent context is social interaction, for example, including bystanders in the VR experience \cite{gugenheimerShareVREnablingCoLocated2017, gugenheimerFaceDisplayAsymmetricMultiUser2018} or managing spaces between VR and non-VR users \cite{yangShareSpaceFacilitatingShared2018}. Recent HMDs such as the Oculus Quest are mobile and standalone systems. They are easier to be carried around to new environments.
Prior research has explored using VR in different forms of transportation (e.g., planes \cite{williamsonPlaneVRSocialAcceptability2019}, cars \cite{hockCarVREnablingInCar2017, mcgillAmPassengerHow2017}, and others \cite{mcgillVirtualRealityPassenger2019, mcgillChallengesPassengerUse2020}).

In this work, we focus on the confined space, where a user cannot fully extend one's limbs. This includes cars, planes, and other forms of public transportation. An example would be maintaining presence and productivity in VR while the user is confined in the rear seat of a car \cite{liRearSeatProductivityVirtual2021}. In these cases, users are usually restricted to a small physical space (e.g., car and shuttle) or surrounded by passengers (e.g., train), where whole-body motion is almost impossible. We explore how to keep presence and enjoyment in VR experiences within the context of the confined space by using beyond-real interaction---mapping finger motions onto virtual arms. 

\subsection{Improving Ergonomics for VR interaction with Hands and Arms}
\paragraph{\textbf{Reality-Based and Illusory Interaction}}
Directly using reality-based interaction (e.g., manipulation using hand tracking) benefits from our real-life experiences. Compared to ray-casting with controllers (or other metaphors), manipulating with hands is more natural and intuitive. Hand tracking benefits from the Degrees of Freedom (DoF) in manipulating virtual objects. Therefore, it becomes an ideal input for VR. On the contrary, ray-casting needs additional extensions to allow for rotation and retraction (e.g., controlling the depth of the ray \cite{laviola3DUserInterfaces2017}) because the manipulation of ray-casting is not hand-centered. Additionally, simulating whole-body movements we have in the real-world experience also enhances presence \cite{slaterInfluenceBodyMovement1998, usohWalkingWalkinginplaceFlying1999} and engagement \cite{isbisterGuidelinesDesignMovementBased2015} in VR.

Although reality-based interaction is natural, they sometimes cause fatigue and require physical space for whole-body motions. Prior works leverage visual dominance over proprioceptive cues \cite{burnsHandSlowerEye2005} to create interaction techniques that reduce fatigue without breaking the Sense of Embodiment (SoE) and the reality-based metaphor (i.e., illusory interaction in \cite{abtahiBeingRealSensorimotor2022}). SoE has three working constructs: self-location, sense of agency (being able to act and control), and body ownership \cite{kilteniSenseEmbodimentVirtual2012}. Recent research investigates how to reduce the physical path length and fatigue of limbs within the maximum arm's reach while preserving VBO. Examples include redirecting visual targets \cite{montanomurilloErgOErgonomicOptimization2017}, gradually directing the user's hand to a comfortable posture \cite{feuchtnerOwnershiftFacilitatingOverhead2018}, or creating an offset between the physical and virtual hand \cite{liEvaluationCursorOffset2018, wentzelImprovingVirtualReality2020}. However, these techniques still require sufficient physical space for whole-body motions and may be unavailable in the confined space.

% added perception perspective
Using a self-avatar with an input metaphor similar to reality enhances the sense of presence and embodiment in VR. High-fidelity avatars have better distance estimation in the near field {\cite{ebrahimiInvestigatingEffectsAnthropomorphic2018}}, and it takes time for VR users to adapt to the dimensions and capabilities of self-avatars in VR {\cite{dayExaminingEffectsAltered2019}}. Although using a self-avatar has been state-of-the-art, recent research explores avatars beyond reality. Won et al. {\cite{wonHomuncularFlexibilityVirtual2015}} studied how participants control an extended third arm controlled by feet in VR and showed they could learn and achieve higher task performance. In a recent study, Kato et al. {\cite{kondoReassociationBodyParts2020}} showed using visual-motor synchrony can enable small VBO when mapping the right-hand thumb motion to the left virtual arm in VR. FingerMapper is inspired by mapping smaller finger motions to the virtual arm. Our approach can be an alternative when interacting in a confined space but the user still wants to preserve SoE.

\paragraph{\textbf{Beyond-Real Interaction as a Solution for the Confined Space}}
VR interactions can go further beyond replication of reality \cite{abtahiBeingRealSensorimotor2022}. Previous research explores the beyond-real concept to enable distant interaction \cite{poupyrevGogoInteractionTechnique1996, feuchtnerExtendingBodyInteraction2017} and object manipulation \cite{freesPRISMInteractionEnhancing2007}. More recent examples aim at reducing physical movements also fatigue for tapping \cite{mcintoshIterativelyAdaptingAvatars2020a} and target selection tasks \cite{schjerlundNinjaHandsUsing2021a}. \FingerMapper\ reduces physical path length and fatigue within the maximum arm's reach of the VR user. Inspired by the beyond-real concept, we focus on mapping small-scale finger motions onto the virtual arms for an energy-efficient interaction, reducing physical movements to avoid fatigue for VR and collisions inside confined spaces. In addition, we are inspired by the small-scale mid-air gestures like Gunslinger \cite{liuGunslingerSubtleArmsdown2015}, which redirects a ray-casting interaction from an energy-efficient position at the user's hip towards a pointing gesture. Dukes et al. {\cite{dukesPunchingDucksPoststroke2013}} design a novel 3D interaction that scales the stroke survivor's impaired arm movements into full movements in VR to provide expert action observation for post-stroke neurorehabilitation. Although these approaches are not resolving physical constraints, \mbox{\FingerMapper} shares the concept of mapping smaller motions to more physically demanding interactions. When the physical space is limited, FingerMapper maps smaller finger motions to the virtual arm to enable a whole-body experience in VR. By reducing physical movements, we expect FingerMapper to have fewer collisions and enhance the VR user's perceived safety.
%%%%%%%%%%%%%%%%%%%%%%%%%%%%%%%%%%%%%%%%%%%%%%%%%%%%%%%%%%%

\section{Interaction Design}
Our design is to enable virtual hands and arms for VR interaction in confined spaces. Hand tracking and ray-casting are unavailable because the former may obstruct the surroundings, and the latter only provides a pointing metaphor. \FingerMapper\ aims to have the same functionality as hand tracking---having arm motion for object interaction---but less physical motion and fatigue. 

\subsection{Mapping Functions}
\begin{figure*}[t]
 \centering % avoid the use of \begin{center}...\end{center} and use \centering instead (more compact)
 \includegraphics[width=\linewidth]{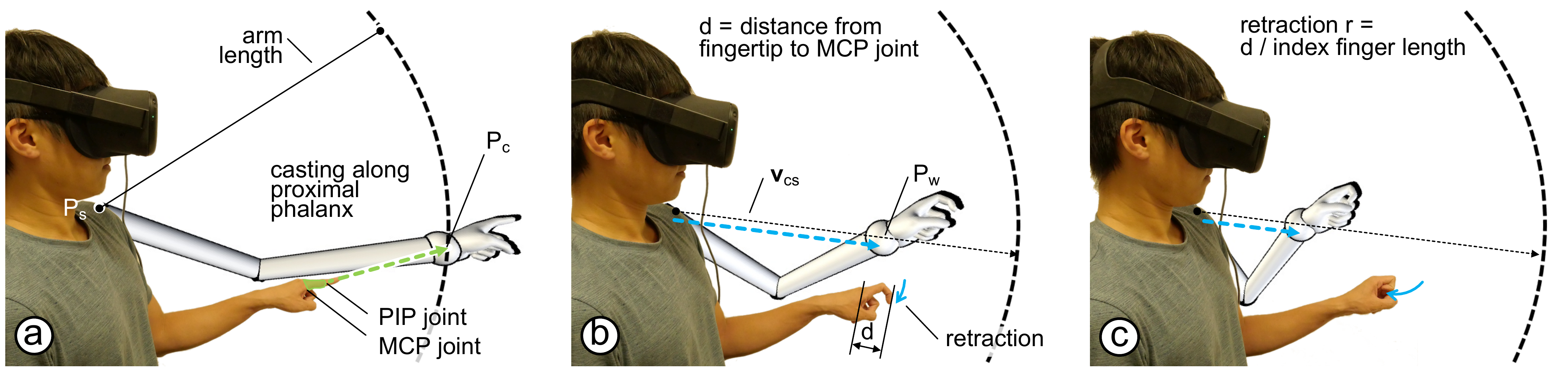}
 \caption{(a) \textit{\Attach}\ casts a ray along the proximal phalanx to the maximum arm reach sphere.  (b)(c) When user bends their finger, we compute a retraction fraction with $d$ to retract the virtual wrist towards shoulder.}
 \Description{Figure 2a depicts the implementation of Attach mapping. We casts a ray along the proximal phalanx to a sphere, whose origin is attached on the user’s shoulder and its radius is the user’s maximum arm length. The wrist of a hand model locates at this casting point. Figures 2b and 2c show the user retracts the virtual arm toward the shoulder by bending their index finger.}
\label{fig:technique-attach}
\end{figure*}

Two mapping functions (\mbox{\textit{Attach}} and \mbox{\textit{Direct}}) for controlling the virtual arm using finger motions were implemented. The maximum arm reach was represented by a sphere having an origin at the shoulder ($P_s$) and a radius of the user's arm length ($l_{arm}$, Figure {\ref{fig:technique-attach}}a). We set $l_{arm}$ (from shoulder to wrist) as 60 cm according to the average arm length  {\cite{pheasantBodyspaceAnthropometryErgonomics2005}}. The goal was to use hand tracking data (e.g., position and rotation of finger joints and wrist) to calculate the virtual wrist position ($P_w$) and create virtual arm movements that can cover the VR user's maximum arm reach.

\paragraph{\textbf{Attach: Casting the virtual wrist onto the surface of the maximum arm reach.}}
The \textit{Attach} mapping design was based on homuncular flexibility in that humans are flexible at learning a different relationship between tracked and rendered motion \cite{wonHomuncularFlexibilityVirtual2015}. First, we cast a ray along the direction from the metacarpophalangeal (MCP) joint to the proximal interphalangeal (PIP) joint. This ray reaches the surface of the maximum arm reach sphere (Figure \ref{fig:technique-attach}a). We chose the proximal phalanx since it is less intervened when bending the finger inwards and called this intersection point the casting position ($P_{c}$). The wrist of a hand model locates at $P_{c}$. Here, the user's index finger is in the pointing gesture, so the virtual arm is also fully stretched. They can move their proximal phalanx of the index finger or their wrist to cast $P_{c}$ on the surface of the maximum arm reach, changing the virtual arm's direction and position.

Next, to reach positions inside the maximum arm reach sphere, we retracted the virtual wrist along the direction between the casting point ($P_c$) and the shoulder ($P_s$). A retraction fraction $r$ was the distance between the fingertip and MCP joint ($d$) divided by the length of the index finger. As the user bends their index finger, $r$ decreases and the virtual wrist moves closer to the shoulder (Figure \ref{fig:technique-attach}bc). Because the virtual wrist should not pass through the VR user's shoulder, the value of $r$ ranges from 0.15 to 1. Therefore, the virtual wrist position ($P_{w}$) was the retraction amount multiplied by the arm length and the unit vector from the shoulder position to the casting point (Equation \mbox{\ref{eqn:attachPosRetract}}). Note that $P_c$ and $P_w$ are identical when $r$ equals one. While the user retracts their index finger, $P_w$ moves toward the user, along the direction from $P_c$ to $P_s$.
\begin{equation}
  \label{eqn:attachPosRetract}
  P_{w} = P_{s} + r \cdot l_{arm} \cdot \hat{\mathbf{v}}_{cs},\ where\ r\ is\ 0.15 < r < 1\ and\ \mathbf{v}_{cs} = P_{c} - P_{s}
\end{equation}

% \hl{Finally, the user rotates the virtual arm as a whole by moving their index finger. 
Finally, because \mbox{\Attach} only calculates $P_w$, we compute the direction of the virtual upper and lower arm by solving inverse kinematics, in which $P_{w}$ is the target, and $P_s$ is the origin. We constrained the virtual upper and lower arm movements according to human arm anatomy. The idea is to have more natural virtual arm movements. We applied 1€ Filter \cite{casiezFilterSimpleSpeedbased2012} to smoothen virtual arm movement.

\begin{figure}[t]
 \centering % avoid the use of \begin{center}...\end{center} and use \centering instead (more compact)
 \includegraphics[width=\columnwidth]{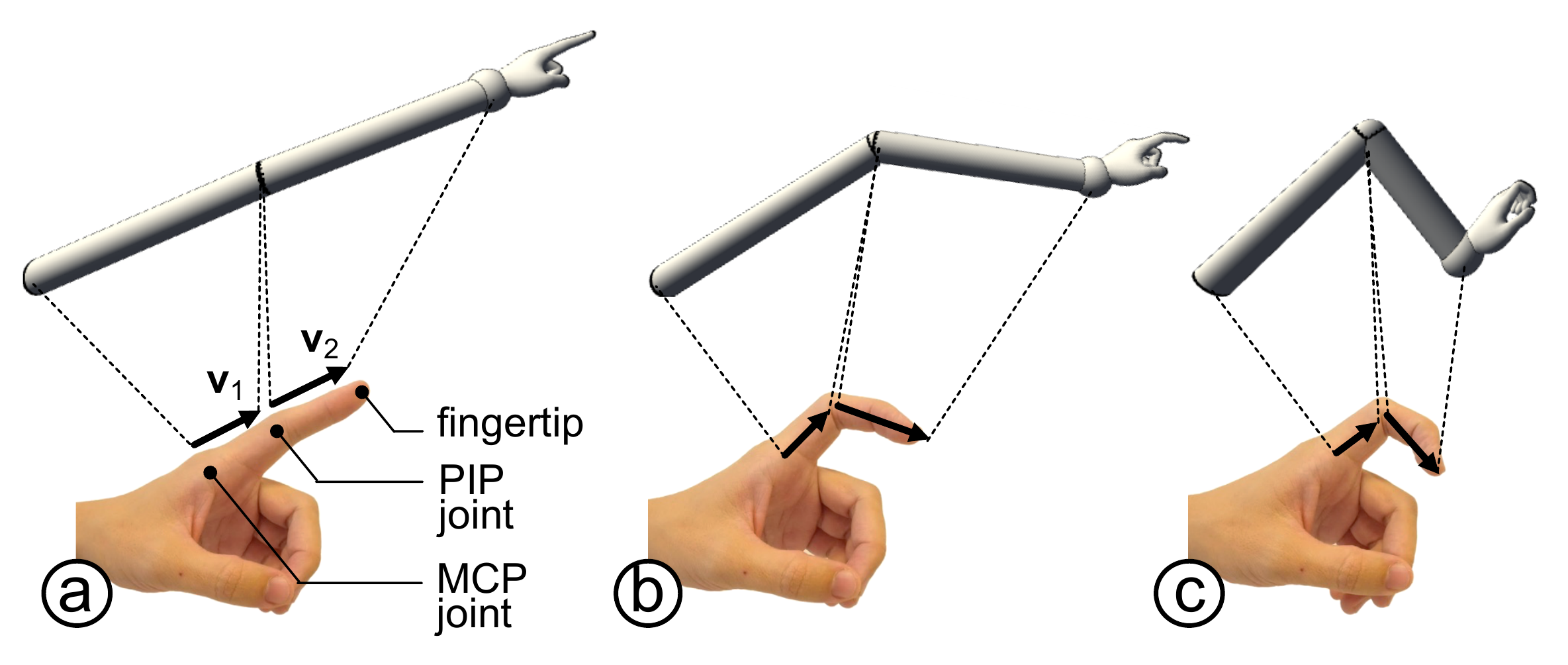}
 \caption{(a) \textit{\Direct}\ maps two vectors, the MCP-PIP and the PIP-fingertip to the virtual upper and lower arms. When the index finger is fully stretched, the virtual arm follows the direction of two vectors and becomes fully extended, (b) bent, and (c) retracted closer to the body.}
 \Description{Figure 3a shows the direct mapping in which the virtual arm angle follows the direction of two vectors on the index finger. Figures 3b and 3c illustrate when the user bends their index finger, the virtual arm moves accordingly and retracts closer to the shoulder.}
\label{fig:technique-direct}
\end{figure}

\paragraph{\textbf{Direct: Re-associating finger segments to virtual arms.}}
The \textit{\Direct}\ mapping function is inspired by body re-association and the notion that providing visuomotor synchrony could induce a portion of VBO \cite{kondoReassociationBodyParts2020}. By using the similarity of index finger joints and virtual arm joints, users control the virtual arm movements by bending their index fingers. The idea is to use this similarity of posture between finger and arm to give the user a better understanding of agency and VBO.

Two vectors, MCP to PIP joint (${\mathbf{v}_1}$) and PIP joint to fingertip (${\mathbf{v}_2}$), are mapped onto the upper and lower virtual arms (Figure \ref{fig:technique-direct}a). When the user's index finger is straight, the virtual arm also remains in the same direction and reaches the maximum arm reach. While the user bends their index finger (Figure \ref{fig:technique-direct}b), the virtual arm follows the direction of two vectors on the index finger and bends inwards simultaneously. Equation \mbox{\ref{eqn:directPosRetract}} shows the virtual wrist position ($P_{w}$) is the sum of ${\mathbf{v}_1}$ and ${\mathbf{v}_2}$ from the shoulder position ($P_{s}$). We scaled up by multiplying half of the arm length ($l_{arm}$).

Figure \ref{fig:technique-direct}c shows the arm length decreases by a fraction $r$ as the retraction amount of virtual arm length while bending the finger. Here, $r$ equals the distance between the fingertip and MCP joint divided by the length of the index finger, ranging from 0.15 to 1. Because the virtual upper and lower arm lengths are equal, the new length equals $r$ multiplied by half of the arm length ($l_{arm}/2$). The virtual wrist position ($P_{w}$) becomes the sum of two unit vectors of the index finger and multiplies the retracted arm length from $P_s$ (equation \mbox{\ref{eqn:directPosRetract}}). \hl{The user can control the virtual arm direction by moving their fingers, and the whole virtual arm rotates when the user rotates their wrist along the direction of the index finger. The difference between \mbox{\Attach} is that the arm direction follows the index finger's angle.} Our implementation applied 1€ Filter \cite{casiezFilterSimpleSpeedbased2012} to hand tracking data for smoothening virtual arm motions.

\begin{equation}
  \label{eqn:directPosRetract}
  P_{w} = P_{s} + r \cdot (l_{arm}/2) \cdot (\hat{\mathbf{v}}_1 + \hat{\mathbf{v}}_2),\ where\ r\ is\ 0.15 < r < 1
\end{equation}

\subsection{Spatial Extension}
% why we need this part
While \mbox{\Attach} and \mbox{\Direct} enable virtual arm movements in three dimensions within the maximum arm's reach, we found that the sense of agency over the virtual arm breaks when moving the wrist without the corresponding virtual wrist movement. We observed this effect during the early prototyping process of FingerMapper. The sense of agency is a construct of being able to control and manipulate, and it is sensitive to any temporal discrepancies between the execution of a self-generated movement and visual feedback \mbox{\cite{kilteniSenseEmbodimentVirtual2012}}. Therefore, we developed a spatial extension function that changes the virtual arm length to create virtual wrist movements, maintaining the sense of agency and overall SoE while using FingerMapper. We applied this function to both Attach and Direct.

The spatial extension was implemented with a similar approach to the Go-Go technique \mbox{\cite{poupyrevGogoInteractionTechnique1996}}. The Go-Go technique is designed for distant interaction in the virtual environment (from mid-range to far-range) and requires enough space for whole-body movement (need to extend arm). On the contrary, \mbox{\FingerMapper} enables virtual arm interaction in the ``close and mid-range'' space (within arm span) using mainly finger mapping. The two techniques have distinct design objectives. In FingerMapper, users can move only their fingers to reach a position in space and their wrists to support their motion. The observed behavior often consists of both. Depending on the context of usage, one of them would become more dominant but never exclude the other. FingerMapper only applies spatial extension to preserve the sense of agency while interacting within close and mid-range. The part of Go-Go is only triggered during larger motions, and the close-range interaction is supported only by FingerMapper. The Go-Go technique in a confined space would become hand tracking in our scenario.

\begin{figure}[t]
 \centering % avoid the use of \begin{center}...\end{center} and use \centering instead (more compact)
 \includegraphics[width=\columnwidth]{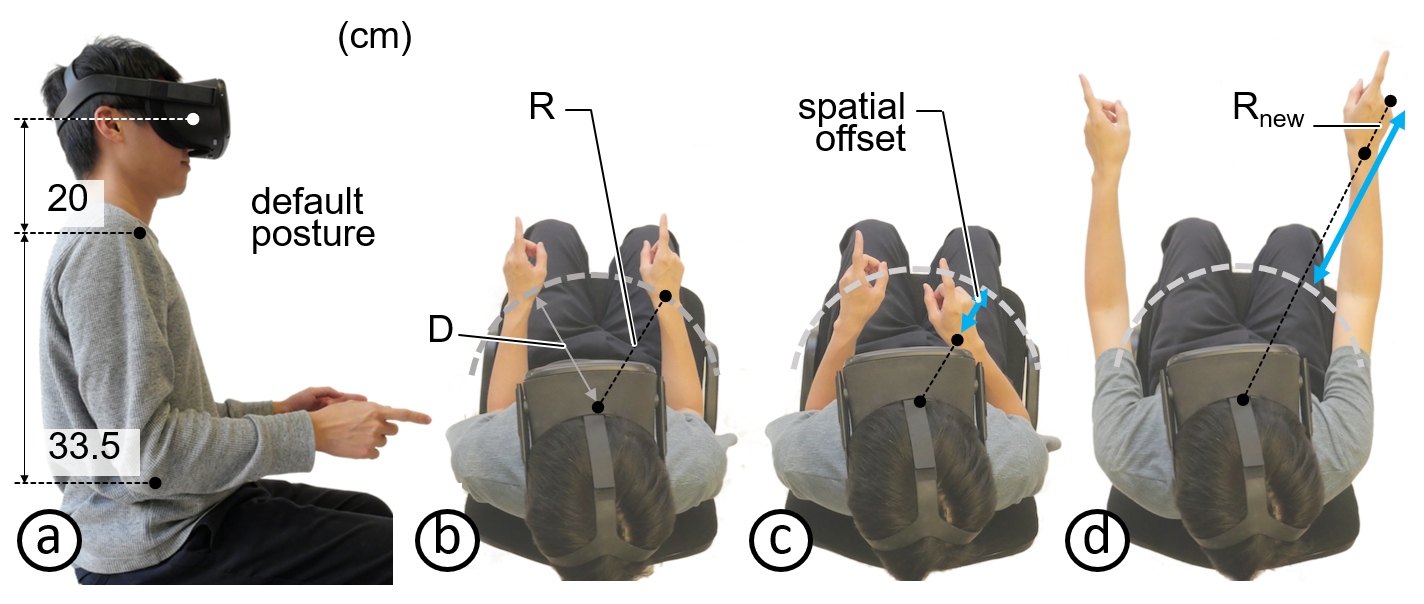}
 \caption{(a) The default posture of \FingerMapper. (b) $R$ is the projected hand-to-chest distance, and $D$ is the default distance for calculating spatial offset. (c) When user moves their hands in the range of $D$, we reduce the virtual arm length by the spatial offset (the blue arrow). (d) Otherwise, we extend $R$ to $R_{new}$ through our extension function and increase the virtual arm length.}
 \Description{Figure 4a shows the default posture where a VR user sits in a confined space and places elbows to their waist. Figures 4b, 4c, and 4d depict spatial extension function for preserving the sense of agency while using Attach and Direct. When the user's wrist exceeds distance D (18 cm), we extend the virtual arm length to match the wrist movement.}
\label{fig:technique-ext}
\end{figure}

% implementation (show the design decision)
Figure \ref{fig:technique-ext}a shows the default posture that the user takes a sitting position and fixes their elbows to both lateral sides of the waist. The default posture simulates using VR in a confined and restricted interaction space. The sitting shoulder height is 20 cm below the eye, and the sitting elbow height is 33.5 cm below the shoulder, approximated using the average body dimensions \cite{pheasantBodyspaceAnthropometryErgonomics2005}. The chest position is 37 cm below the HMD, and the chest height is in the middle between the sitting shoulder and elbow height. We calculate the projected distance from the physical wrist to the chest ($R$), as shown in Figure \ref{fig:technique-ext}b. This distance is projected onto the plane that always has its normal vector pointing upwards from the user's head since we need the forward and backward spatial offset. The projected distance ($D$) of default posture is the criterion of spatial extension. We set the value of $D$ as 18 cm through our first experimentation. As shown in Figure \ref{fig:technique-ext}c, when the user's hand is within the range of $D$, The spatial offset is $D - R$. When the user extends their hand out of the range of $D$ (Figure \ref{fig:technique-ext}d), we compute $R_{new}$ using the Go-Go function shown in equation \ref{eqn:ext}. The spatial offset then becomes $R_{new} - D$. These design decisions of the spatial extension were made during the implementation process within the group of authors. The offset is added to the radius of the shoulder sphere of \Attach. Therefore, the user can perceive the arm extending or shortening when they move their hand forward and backward. For \Direct, we increase or decrease the arm length by this offset. 

\begin{equation}
  \label{eqn:ext}
R_{new} =\left\{ \begin{array}{ll}
      R              & {if\ R\ <\ D} \\
      R + k(R - D)^2 & {otherwise},\ k\ =\ 0.6
    \end{array}
    \right.
\end{equation}

\subsection{Finger Motion}
In an early prototype, we experimented with a rigid hand model as the virtual hand that either points or grabs. However, this static representation felt more awkward since the virtual hand model did not resemble the finger motions we have in reality. To increase the sense of agency and VBO, we map the tracking information of physical fingers to virtual fingers. Although index fingers are occupied for positioning the virtual arm in space, the current implementation still has more precise finger movement at this new location and performs simple manipulations. 

\begin{figure}[t]
 \centering % avoid the use of \begin{center}...\end{center} and use \centering instead (more compact)
 \includegraphics[width=\columnwidth]{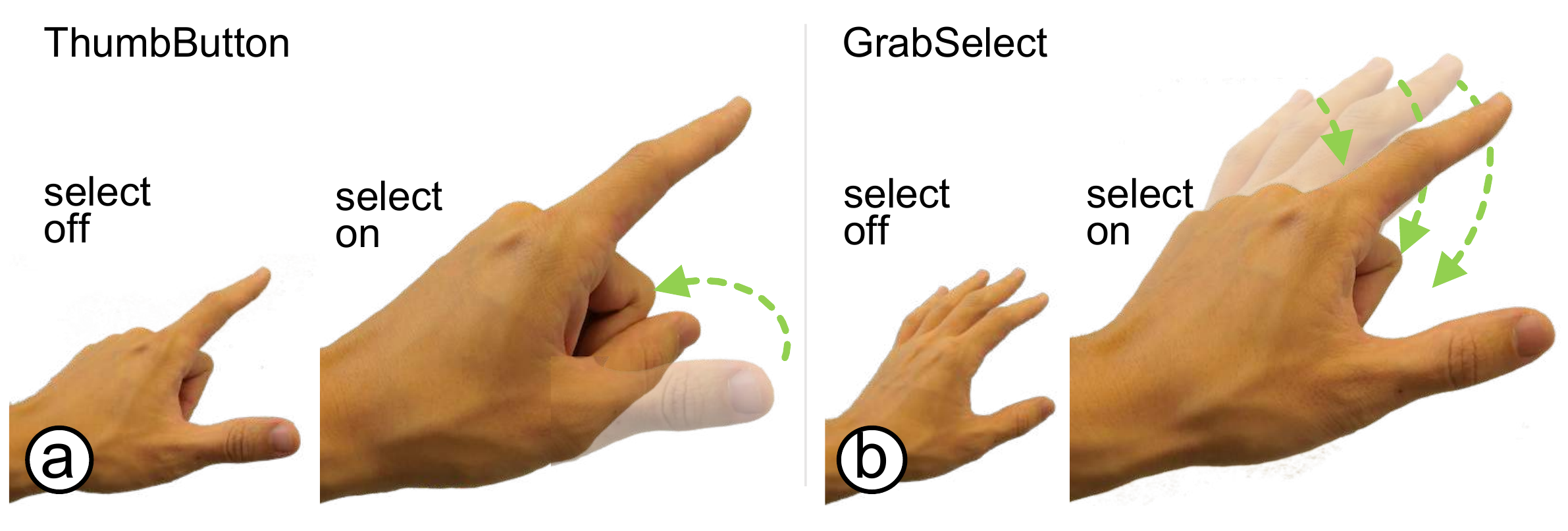}
 \caption{(a) \textit{ThumbButton} uses the distance between the thumb fingertip and middle finger knuckle to select. (b) \textit{GrabSelect} uses the metaphor of holding an object. }
 \Description{Figures 5a and 5b show two selection techniques for FingerMapper, ThumbButton, and GrabSelect. ThumbButton uses the distance between the thumb fingertip and middle finger knuckle as a threshold to trigger selection. GrabSelect is a metaphor like grabbing an object with a selection threshold calculating the average distance from the fingertips of the middle, ring, and pinky fingers to the wrist.}
\label{fig:selection}
\end{figure}

\subsection{Selection}
Selection is essential in VR interaction. Despite having the actual finger tracking data attached to the virtual hand, the current implementation partially impedes the ability to select objects with a pinch since bending the index finger also moves the virtual arm. To overcome this limitation, we added functionality based on two hand gestures (\emph{ThumbButton} and \emph{GrabSelect}) to enable selection with \FingerMapper.  

In \emph{ThumbButton}, the user moves their thumb to touch the knuckle of the middle finger, similar to pressing a button (Figure \ref{fig:selection}a). We track the distance from the thumb fingertip to the middle finger knuckle (PIP joint). The user triggers selection once detecting the thumb fingertip contacts the middle finger knuckle. Figure \ref{fig:selection}b shows \emph{GrabSelect}, where the user holds their hand with the middle, ring, and pinky fingers at the same time. We compute an average distance from the middle, ring, and pinky fingertips to the wrist, leveraging the metaphor of grabbing an object to trigger selection. 

We tested both trigger mechanisms within the group of authors during the implementation process of FingerMapper to make design decisions. While performing the motion in GrabSelect feels more natural and closer to grabbing an object, tucking in the middle, ring, and pinky finger results in a higher distortion of the index finger position. \hl{These joint movements with the index finger are because the fingers are controlled by the same muscle group \mbox{\cite{schwarzAnatomyMechanicsHuman1955}}. This distortion causes the Heisenberg effect of spatial interaction \mbox{\cite{wolfUnderstandingHeisenbergEffect2020}}---a user points or selects virtual content using spatially tracked input devices, and a discrete input such as a button press disturbs the tracking position, resulting in a different selection point. Here, GrabSelect disturbed the index finger position, so this effect exists.} Therefore, We selected ThumbButton since the muscle group involved in the thumb motion is independent of the index finger, resulting in less disturbance of the selection.

\subsection{Hardware Requirement}
Our current setup requires 6-DoF tracking on the wrist, joints, and fingertips. We use Oculus Quest hand tracking to acquire all the necessary position and rotation data. The concept is also compatible with other VR systems providing similar hand-tracking data.

%%%%%%%%%%%%%%%%%%%%%%%%%%%%%%%%%%%%%%%%%%%%%%%%%%%%%%%%%%%
\section{Applications}
\hl{We highlight three usage scenarios of \mbox{\FingerMapper} with different interactions (e.g., object interaction, arm swinging motion, and locomotion). The goal is to show that when the physical environment is unavailable for whole-body movement, the user can switch to \mbox{\FingerMapper} as an input metaphor for confined space. Three applications were originally designed for whole-body interaction. We only replace the regular hand tracking (or controllers) with \mbox{\FingerMapper} to show VR users can interact in a confined space using \mbox{\FingerMapper} in different scenarios.}

\subsection{\FingerGrabber}
\label{sec:FG}
% ref: https://www.mariowiki.com/File:SMP_image9.png

%Inspiration of the Game + what is the goal and how to achieve it
\FingerGrabber\ is an application inspired by VR sandbox games like Job Simulator. The interaction in this type of game is often selecting, grabbing, and manipulating any object in the virtual environment. This interaction style can represent any generic selection and manipulation task a user has to conduct in VR. The virtual objects in these VR applications are usually scattered within the arm span of the user. Here, we argue that \FingerMapper\ could be a good solution by reducing physical movement to enable the gaming experience in confined spaces.

\begin{figure}[t]
 \centering
 \includegraphics[width=\columnwidth]{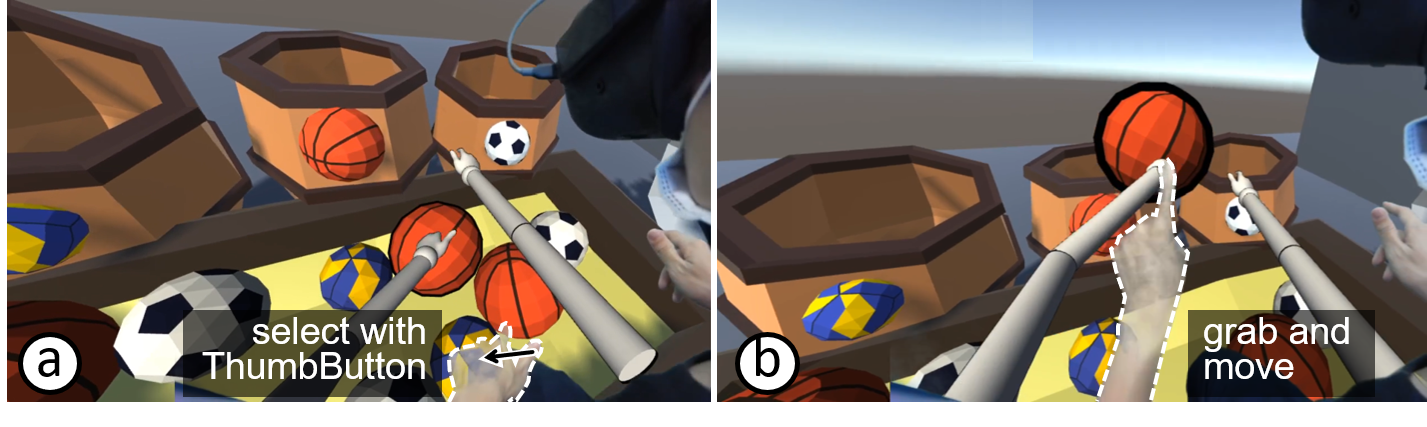}
 \caption{(a) FingerGrabber is a sandbox that the player selects and manipulates objects surrounding them (e.g., in our application, basketball, volleyball, and football). The player has to select the ball using \Attach\ mapping and put (b) the ball into the baskets in VR.}
 \Description{Figure 6a illustrates FingerGrabber, where the VR user selects three types of balls (e.g., basketball, volleyball, and football) and puts them into the corresponding bucket. Figure 6b shows a user manipulates an object with FingerMapper, enabling object interaction in VR.}
 \label{fig:app-grabbingroom}
\end{figure}

Since such games often strive for a well-written narrative, \FingerGrabber\ added game mechanics to have a more engaging experience. The user is in front of a pit that is successively filled with different types of balls (e.g., basketball, volleyball, and football shown in Figure \ref{fig:app-grabbingroom}a). Figure \ref{fig:app-grabbingroom}b shows that the user grabs each ball and place it in one of the three corresponding baskets. We implement the game using \Attach\ mapping and \emph{ThumbButton}, allowing the player to conduct the full range of interactions necessary for this type of game, i.e., select, grab, and manipulate.  

\subsection{\FingerSaber}
\begin{figure}[t]
 \centering % avoid the use of \begin{center}...\end{center} and use \centering instead (more compact)
 \includegraphics[width=\columnwidth]{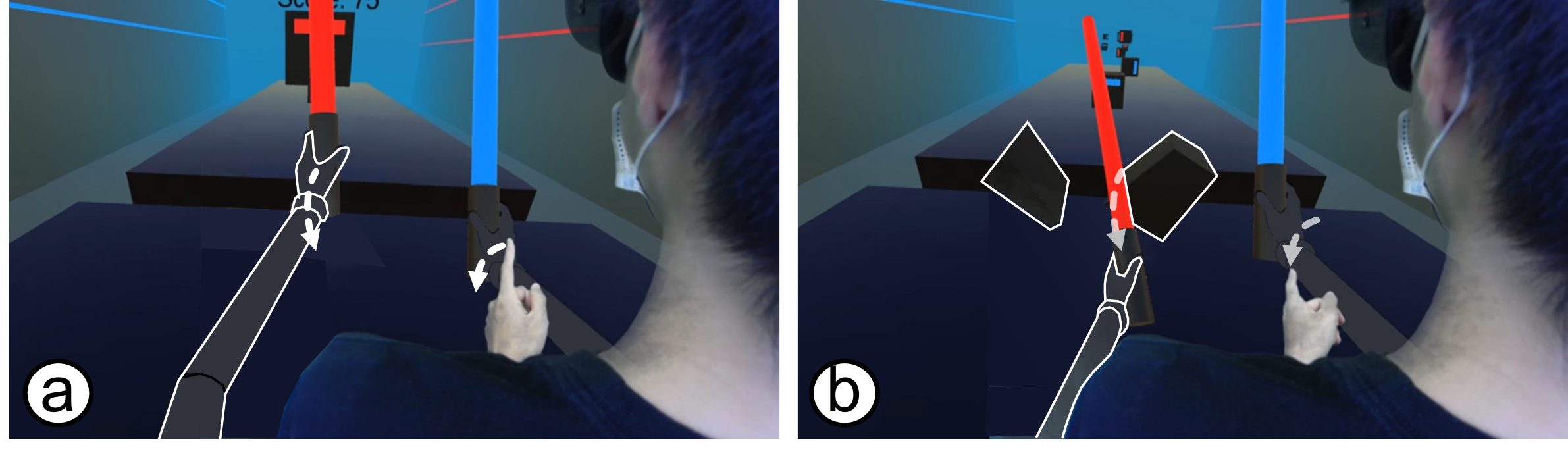}
 \caption{(a) We use \Attach\ mapping to control virtual arms with lightsabers for swinging motion. (b) The user has to slice cubes in the correct direction and corresponding color.}
 \Description{Figure 7a depicts FingerSaber, in which the VR user moves the index finger to replace arm-swinging motions with FingerMapper. Figure 7b shows the goal is to slice every incoming cube in the indicated order.}
\label{fig:app-fingersaber}
\end{figure}

%Inspiration of the Game + what is the goal and how to achieve it
The VR rhythm game, BeatSaber, inspired us to use fingers to perform the arm swing motion (Figure \ref{fig:app-fingersaber}a). In BeatSaber, different colored cubes move toward the user and follow the beats of the music. The user's goal is to slice every cube in the direction indicated on them. The original concept of BeatSaber revolves around having a whole-body experience with large arm movements and sometimes even moving their whole body. 
%How to Interact
To enable \FingerSaber, we chose the \Attach\ mapping and positioned two virtual lightsabers in the user's virtual hand. Since no selection is needed, we kept the lightsabers constantly attached to the user's hand without using a grabbing motion. Our technique enables the BeatSaber experience to work using small finger motions instead of large arm motions without modifying the game mechanics.
%What is different to the original
% During the game, we observed that certain actions (e.g., slicing a cube to the right with the right finger) were more difficult then expected. While in BeatSaber a user would position the body accordingly by raising the elbow and rotate the hand, in \FingerSaber\ the same motion is more uncomfortable due to the DoF of the hand and wrist. Nevertheless, 

\begin{figure}[t]
  \centering
  \includegraphics[width=\columnwidth]{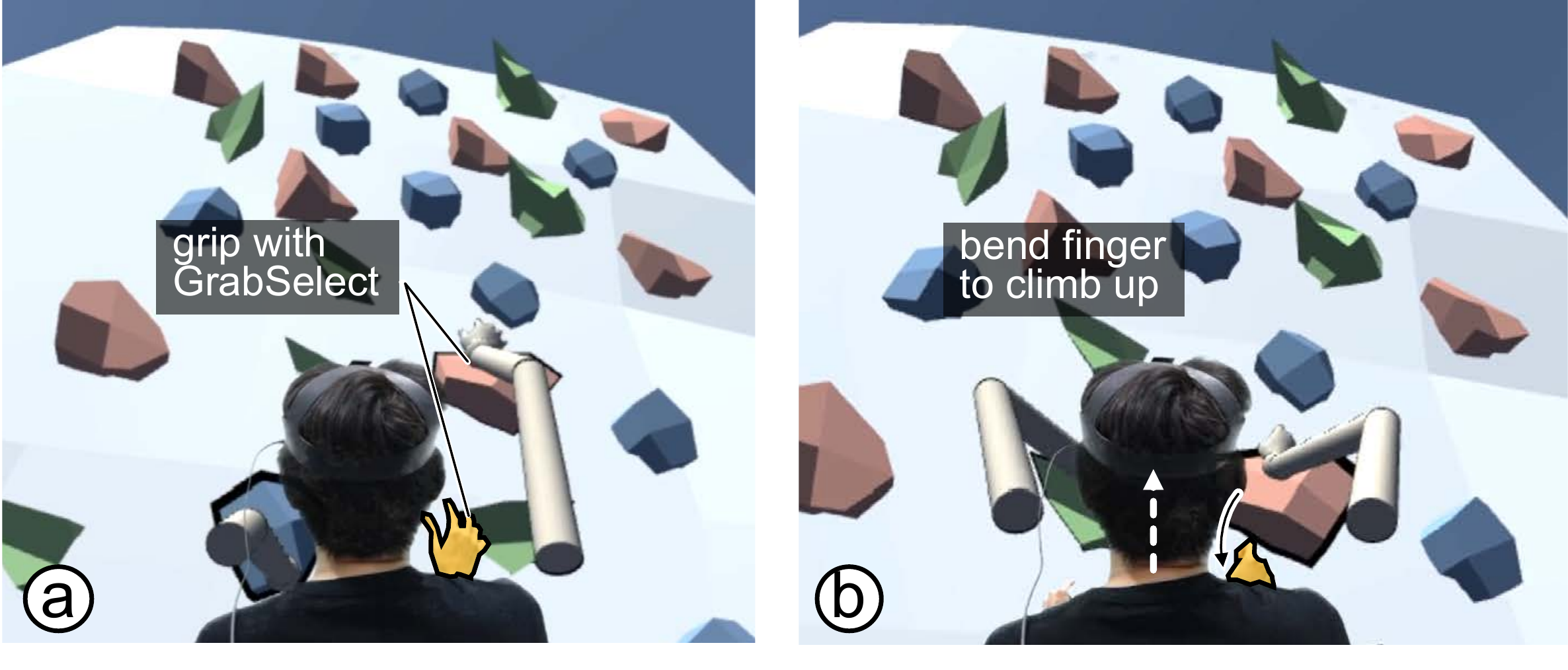}
  \caption{(a) The user selects a grip with GrabSelect. (b) While gripping, the user bends the finger to pull the body upwards and create a vertical displacement in VR.}
  \Description{Figure 8a shows FingerClimber, where a VR user performs locomotion with FingerMapper. Figure 8b depicts that the VR user climbs up by selecting different climbing grips with our technique.}
  \label{fig:app-climb}
\end{figure}

\subsection{\FingerClimber}
%Inspiration of the Game + what is the goal and how to achieve it
\FingerClimber\ is inspired by current VR games like The Climb, where the user goes to the top of a mountain using different climbing grips attached to the cliff. In this game, the user usually stretches their whole arm and sometimes the whole body. We used \Direct\ and hoped this mapping would help the user to control the specific shape of the virtual arm. The goal is to avoid users climbing with constantly stretched-out arms. As the selection mechanism, we used \emph{GrabSelect} since it imitates the natural motion of holding onto a climbing grip (Figure \ref{fig:app-climb}a). The user grabs a climbing grip with the virtual arm and pulls themselves up using a downward motion with the finger (Figure \ref{fig:app-climb}b). 

%%%%%%%%%%%%%%%%%%%%%%%%%%%%%%%%%%%%%%%%%%%%%%%%%%%%%%%%%%%%%%%%%%%
\section{Study 1: Target Selection}
We conducted a target selection study to compare {\Attach} and {\Direct} to common inputs (e.g., {\physicalCond} and {\raycastingCond}). The goal was not to outperform the baselines since none of them can achieve our motivation---arm and hand interactions for VR in a confined space. The comparison in this study was to position our prototype techniques between full embodiment ({\physicalCond}) and full efficiency ({\raycastingCond}). We observed how good users could control \FingerMapper\ and how they impacted metrics (e.g., presence, VBO).

% \begin{figure}[t]
%  \centering % avoid the use of \begin{center}...\end{center} and use \centering instead (more compact)
%  \includegraphics[width=\columnwidth]{study-bodydimension.pdf}
% \caption{(a) The study setup. (b) The body dimensions for calibration (arm span, arm length, and index finger length).}
% \label{fig:study-body}
% \end{figure}

\subsection{Apparatus and Setup}
The study was conducted with Oculus Quest 1 either in our lab or remotely at the participants' homes. Participants were seated on a chair. Five remote participants used their devices at home to avoid unnecessary physical contact and crowds. We instructed remote participants to clear a space that did not cause obstructions while using hand tracking. The study application was given to participants as a .apk file. The other participants used the lab setup on the same headset. To treat both setups as similar as possible, remote and lab participants were given the instructions via a website and could ask questions in person or remotely. We disinfected the HMD cushion for lab participants for hygiene concerns.

\subsection{Study Design}

\begin{figure}[t]
 \centering % avoid the use of \begin{center}...\end{center} and use \centering instead (more compact)
 \includegraphics[width=\columnwidth]{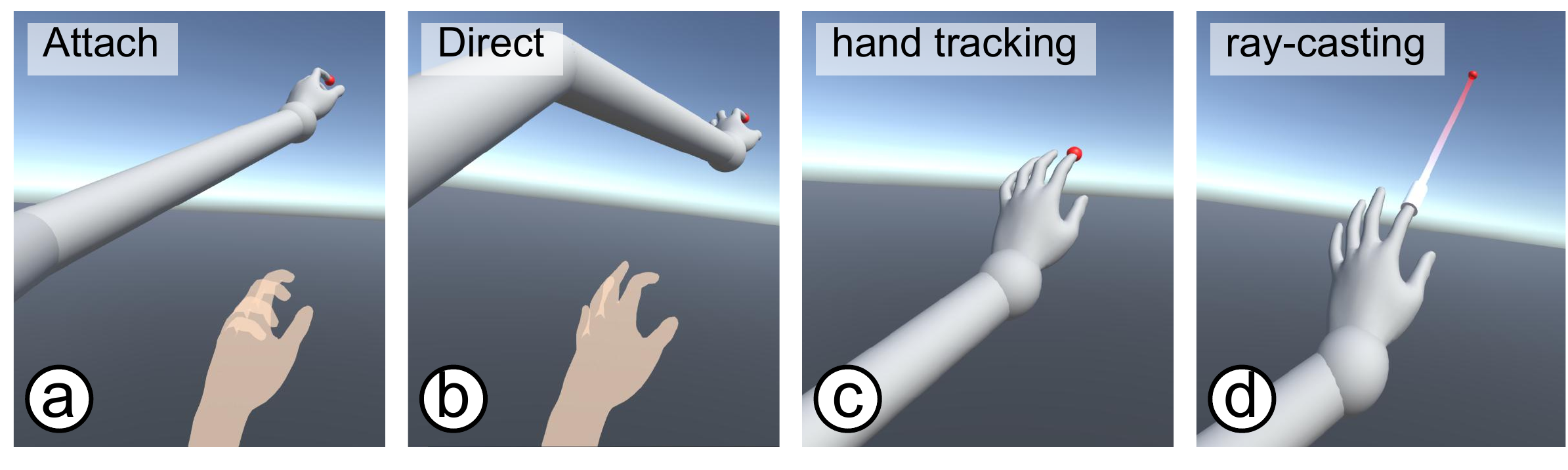}
\caption{We evaluated four \textit{techniques} (a) \Attach, (b) \Direct, (c) \physicalCond, (d) \raycastingCond. The pink hand is for clarifying the real hand position for this figure only. Participants only could see their virtual arms. }
\Description{Figure 9 shows four techniques used in the study, starting with the FingerMapper techniques, Attach and Direct, then two baselines, hand tracking and ray-casting.}
\label{fig:study-hands}
\end{figure}

The study was a within-subject design with one independent variable (\textsc{technique}), including {\Attach}, {\Direct}, {\physicalCond}, and {\raycastingCond} (Figure \ref{fig:study-hands}). We were interested in what degree we could reduce physical movements and fatigue, so we compared with {\physicalCond}. Additionally, {\raycastingCond} was included as a second baseline for understanding efficiency. \hl{Here, we used an implementation similar to \mbox{\cite{bowmanEvaluationTechniquesGrabbing1997}}, where a ray is casting from the fingertip of the index finger (Figure \mbox{\ref{fig:study-hands}d}).} A red pointer was located on the fingertip to visualize the selection point. In {\raycastingCond}, the pointer was on the tip of the ray, indicating the contact point when casting on a target.

We used a world-fixed layout (Figure \ref{fig:study-layout}b) in the target selection task of the previous work \cite{montanomurilloErgOErgonomicOptimization2017, wentzelImprovingVirtualReality2020}. Here, interactable objects were spatially distributed around the user. We argue that this arrangement represents the application scenario with arm and hand interactions in a confined space. The layout dimension was determined by the participant's arm span (\textit{A}). The target layout had 18 spheres on two layers located at the depth of the maximum arm reach (0.44A) and a half of this range (0.22A). The closer layer had 3 $\times$ 4 targets evenly distributed between 0.76A $\times$ 0.4A, and the other layer had 3 $\times$ 2 targets that were placed evenly between 0.25A $\times$ 0.4A from the participant. Targets ranged evenly in the height from 0.4A to 0.8A to simulate a seated interaction space. We calculated the distance between every two targets in the layout and separated them into short-, medium-, and long-range by using the 25th and 50th quantile of distances. Each type of range had an equal amount of tasks in each condition, and they were randomized within 30 tasks of each condition. Each participant had to complete 120 tasks (4 \textsc{techniques} $\times$ 30 tasks). The order of \textit{technique} was counterbalanced by the 4 $\times$ 4 Latin square.

For dependent variables, we measured task time, the physical path length of the used wrist, and the virtual path length of the pointer in VR for each task. The path length data was divided by the target distance of each task as a ratio for comparison. The interaction volume was calculated by recorded physical fingertip trajectory. In the post-test questionnaire, we used NASA-TLX \cite{hartDevelopmentNASATLXTask1988} for the subjective fatigue, the avatar embodiment questionnaire  \cite{gonzalez-francoAvatarEmbodimentStandardized2018} with three subscales (ownership, agency, and location) for VBO, and Igroup Presence Questionnaire (IPQ) \cite{schubertExperiencePresenceFactor2001, regenbrechtRealIllusoryInteractions2002} for presence. Participants responded with a 7-point Likert scale inside VR. Finally, the tracking quality of the Oculus Quest was recorded through every frame during each task for filtering out the data of tracking loss.

\subsection{Procedure and Task}
Participants were introduced to the purpose of our study and signed the consent form. We measured the arm length\footnote{We define arm length as the distance from the shoulder to the wrist. Please see the figure in the supplementary material.}, index finger length, and arm span for calibration. After participants put on the HMD and entered the application, they had to re-center themselves in VR and calibrate their body dimensions. 

\begin{figure}[t]
  \centering
  \includegraphics[width=\columnwidth]{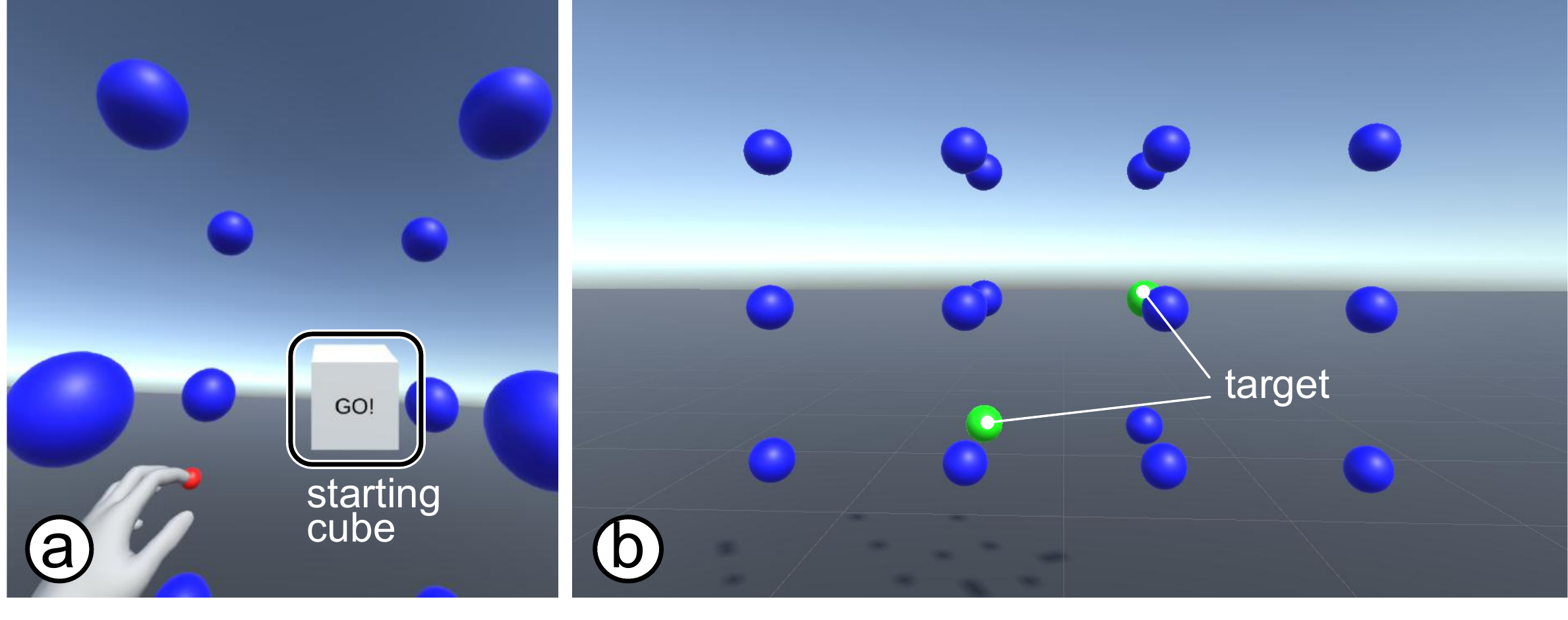}
  \caption{(a) Participants remained default posture before touching the starting cube. (b) The layout with 18 targets. Two green spheres appeared when the participant touched the starting cube.}
  \Description{Figure 10 illustrates the study scene where 18 targets are distributed in a world-fixed layout. Participants touch the starting cube (10a) and reach one of the two green targets to start the task (10b). After selecting one green target, they have to reach the other one as fast as possible.}
  \label{fig:study-layout}
\end{figure}

Participants had to maintain the default posture (Figure \ref{fig:technique-ext}a), placing their elbows on both sides of their waist to simulate a confined space. They touched a white cube in front of them to begin (Figure \ref{fig:study-layout}a), and two randomly selected spheres became green as the targets (Figure \ref{fig:study-layout}b). They were free to choose their right or left hand for a task, but they had to start and finish the task with the same hand. If they used the right hand to select the first target, the second one had to be selected with the right. Using the wrong hand for the second sphere was considered an error. This approach is common in these types of studies \cite{wentzelImprovingVirtualReality2020} and allows users to keep the natural behavior. The task started as the participant touched one of the green targets using the red pointer. This target turned blue, and the participant had to move the pointer to the second target fast and steadily, and this task was marked as finished. A successful task came with a sound effect, and participants returned to the default posture, waiting for a two-second countdown, and continued to the next task by touching the white cube.

% They were also allowed to practice as much as they wanted until they could fully control the interaction.
% Participants had to practice and succeed in at least ten tasks before each condition.
% Each participant was asked to practice with \mbox{\FingerMapper} until they felt confident and comfortable to perform the coming tasks. This should prevent an orientation phase at the beginning of the exposure but resulted in different training duration for each participant. However, since we did not apply any performance metrics in the study but were focusing on subjective measures and collisions, this did not result in a distortion of the data.

\hl{Participants had to practice and succeed in at least ten tasks before starting each condition. They were asked to practice until they felt confident and comfortable performing the task with each technique. This setup should prevent a learning phase at the beginning of the study exposure for each condition but result in different training duration for each participant. The average task count of practicing was 11.1 times ($SD=1.6$), showing that participants had a similar training time. The overall average success rate of practicing was 98\% (\mbox{\Attach}: $M=97.3\%, SD=6.5$, \mbox{\Direct}: $M=100\%, SD=0$, \textit{hand tracking}: $M=97.3\%, SD=5.1$, and \textit{ray-casting}: $M=97.4\%$, $SD=5.1$). These results indicate that participants' mastery of each technique was similar.}
After each condition, they filled out a post-test questionnaire in VR and took a 30-second rest. After finishing all the conditions, we asked for their subjective feedback and preferred interaction if they used VR in a confined space. Each condition took about ten minutes, and the whole study finished in one hour.

\subsection{Participants}
Volunteer participants (n=13) took part in the study via convenience sampling (five females and eight males, age: $M = 26.3, SD = 3.4$). Six participants had experience developing VR or research, and the rest used VR occasionally.

\begin{figure}[t]
  \centering
  \includegraphics[width=\columnwidth]{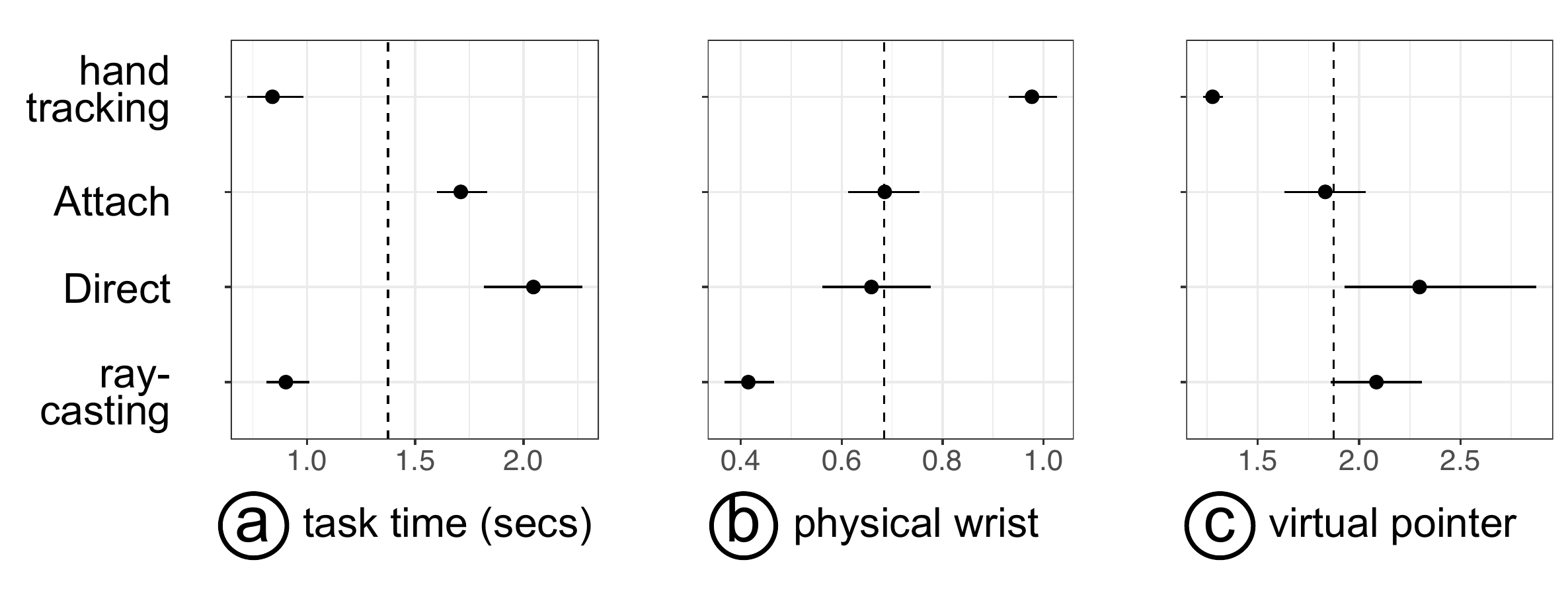}
  \caption{Results of (a) task time in seconds, (b) the path length ratio of physical wrist, and (c) virtual pointer. Path lengths were divided by target distance for each task as a ratio for comparison. Having ratio one means the path length was equal to the target distance in that task. The error bar represents the 95\% CI, and the dotted line shows the average. }
  \Description{Figure 11 shows the results of Study 1. The error bar represents the 95 percent confidence interval, and the dotted line shows the average. Figure 11a depicts the task time in seconds. Attach and Direct had a higher task time (around two seconds) compared to both baselines (about one second) on average. Figure 11b shows the path length of the physical wrist, which was divided by the target distance for each task as a ratio for comparison. Having ratio one means the path length was equal to the target distance in that task. Here, hand tracking has the largest ratio then comes Attach and Direct, and ray-casting has the smallest ratio. This indicates our techniques reduced physical path lengths. Figure 11c illustrates the path length ratio of the virtual pointer. Hand tracking has the lowest ratio then comes to attach, ray-casting, and direct. All pointing metaphors had a higher ratio probably due to overshooting.}
  \label{fig:study-results-quantative}
\end{figure}

\subsection{Results}
The following results were based on 1560 selection tasks. The {\physicalCond} condition had overall more tracking loss than other techniques, probably because \FingerMapper\ and {\raycastingCond} had more static and close-up postures, so they were easier to track. In total, 197 (12.6\%) tasks were removed. The accuracy of each condition was Attach 98.4\%, Direct 99.4\%, hand tracking 99.0\%, and ray-casting 98.1\% (overall 98.7\%). Each technique had a high success rate and showed participants could control the techniques to complete the task. We only considered successful tasks and excluded the task time and path lengths that exceeded three standard deviations from the average in each condition. There were 30 tasks (2.2\%) removed among 1363 well-tracked tasks.

For task time and path length ratios, one-way within-subject ANOVA with Greenhouse-Geisser correction was performed and followed by the pairwise comparison with Bonferroni correction for the post-hoc analysis. We used the Friedman and Wilcoxon test for the post-hoc comparison of the Likert-scale data.

\textbf{Task Time.}
There was a significant difference in \textsc{technique} on the task time ($F_{3, 36} = 90.03, p < .01, \eta^2 = 0.88$), as shown in Figure \ref{fig:study-results-quantative}a. The post-hoc comparison revealed that the mean task time of \Attach\ ($M = 1.71s, SD = 0.24$) and \Direct\ ($M = 2.05s, SD = 0.44$) were significantly higher than \physicalCond\ ($M = 0.84s, SD =  0.25$) and \raycastingCond\ ($M = 0.90s, SD = 0.19$) ($p < .01$). There was no significant difference between \Attach\ and \Direct\ ($p = .09$), neither \physicalCond\ and \raycastingCond\ in the post-hoc analysis ($p = .68$).

\begin{figure}[t]
  \centering 
  \includegraphics[width=\columnwidth]{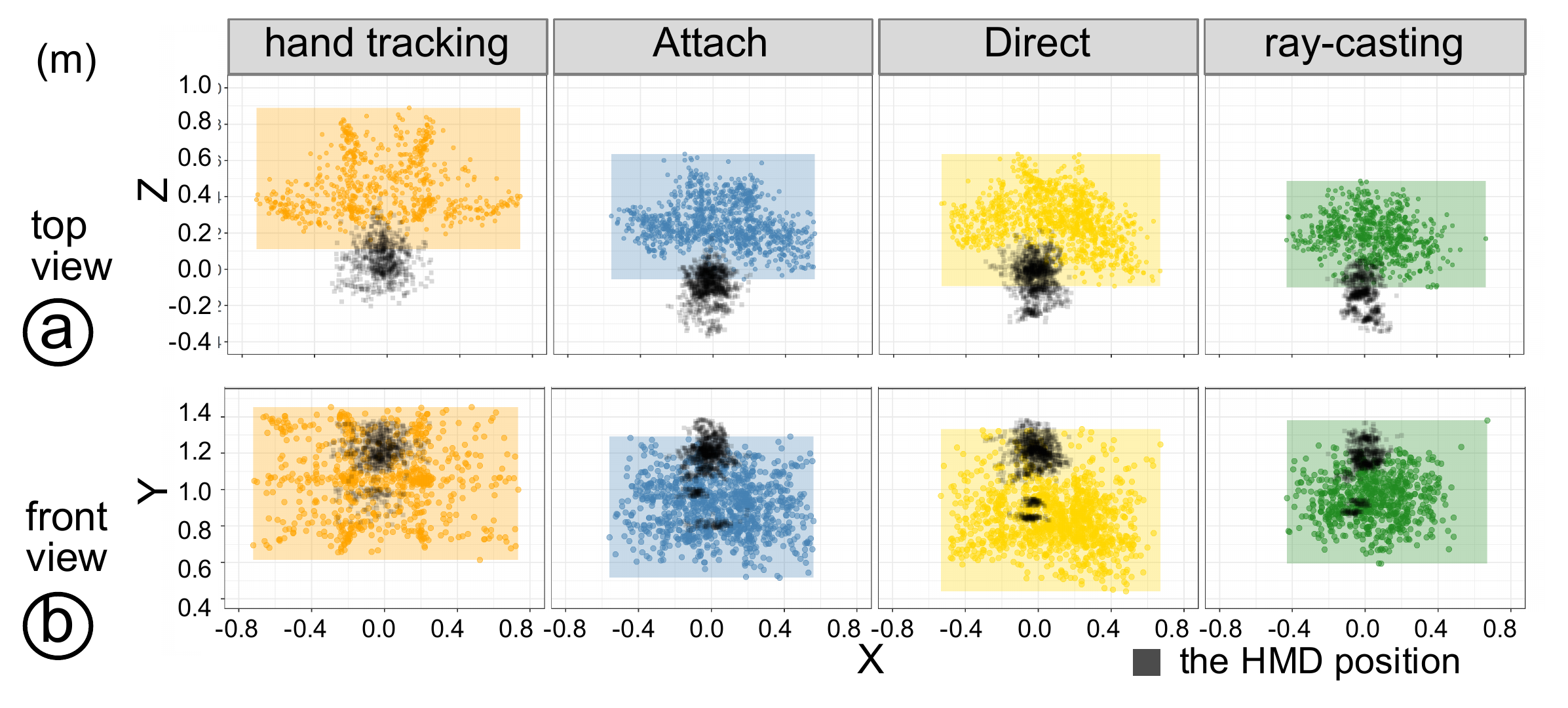}
  \caption{The scatter-plot shows the position (per second) of the index fingertip for all participants. The bounding box represents the area of interaction from the (a) top view and (b) front view. Black squares indicate HMD positions.}
  \Description{Figure 12 shows the position of the index fingertip for all participants. We highlight the bounding box to illustrate the interaction space from the top view (12a) and front view (12b). Black squares indicate HMD positions. We can observe hand tracking had the biggest volume, then came Attach and Direct. Ray-casting had the smallest.}
  \label{fig:study-boundingbox}
\end{figure}

\textbf{Path Length Data.}
All the path length data was divided by the target distance for each task. These metrics represented the relationship between physical movement and virtual distance. Figure \ref{fig:study-results-quantative}b shows the path length ratio of the physical wrist. We found a significant difference in the motion of the physical wrist ($F_{3, 36} = 51.47, p < .01$, $\eta^2 = 0.81$) in \textsc{technique}. In the post-hoc comparison, \raycastingCond\ ($M = 0.42, SD = 0.10$) had a significantly lower path length of the physical wrist compared to the other three conditions ($p < .01$). The results indicated no significant mean difference in the path length of the physical wrist between \Attach\ ($M = 0.69, SD = 0.14$) and \Direct\ ($M = 0.66, SD = 0.20$) ($p = .94$). \textit{Hand tracking} ($M = 0.98, SD = 0.09$) had a significantly higher path length ratio compared to \Attach\ and \Direct\ ($p < .01$).

There was a significant difference in the path length of the virtual pointer ($F_{3, 36} = 11.53, p < .01, \eta^2 = 0.49$). The virtual path length ratio was calculated as the virtual path traveled divided by the target distance (Figure \ref{fig:study-results-quantative}c). \hl{For the virtual pointer of \mbox{\raycastingCond}, we recorded the path length when the virtual pointer was not intersecting with targets. Because the virtual pointer (cursor) was at the end of the ray, there were additional movements on this virtual pointer, resulting in the average ratio was 2.09 for \mbox{\raycastingCond} (Figure \mbox{\ref{fig:study-results-quantative}c)}}. These metrics were an indicator of how much unnecessary movement was done. \textit{Hand tracking} ($M = 1.28, SD = 0.09$) had a significantly lower path length ratio of a virtual pointer than the other three \textsc{techniques} ($p < .01$). Other comparisons revealed no significant mean difference in the virtual path length (Direct-Attach: $p = .054$, Ray-Attach: $p = .51$, Ray-Direct: $p = .65$). \Attach\ ($M = 1.83, SD = 0.40$), \Direct\ ($M = 2.30, SD = 0.89$), and \raycastingCond\ ($M = 2.09, SD = 0.43$) all resulted in significantly more unnecessary motion, indicating overshooting or correction behavior while selecting. Overall, both finger mappings reduced the physical motion but increased the virtual motion compared to \physicalCond. 

\begin{figure*}[t]
  \centering
  \includegraphics[width=\textwidth]{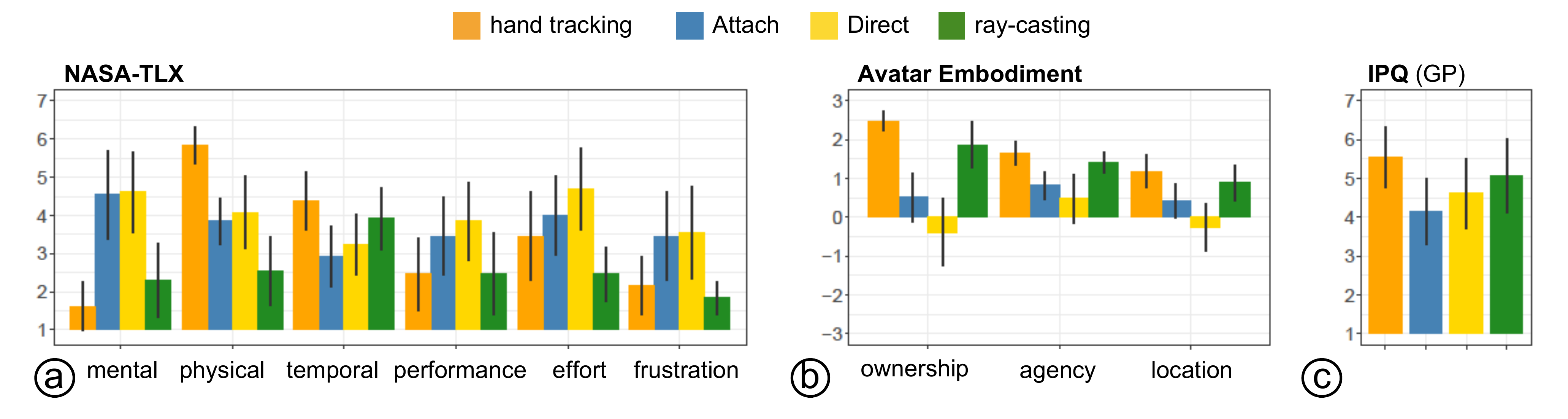}
  \caption{The results of (a) NASA-TLX, (b) ownership, agency, and location subscales from the avatar embodiment questionnaire, (c) the overall score of IPQ. The errorbar represents the 95\% CI.}
  \Description{Figure 13 shows the results of the questionnaire data. The error bar represents the 95 percent confidence interval. Figure 13a illustrates the results from NASA-TLX. For the mental demands, hand tracking and ray-casting received the lowest score. Attach and direct received an average score of 4 out of 7 in mental demanding. This also shows a similar pattern in the performance, effort, and frustration subscale. For the physical demand, hand tracking has the highest score then comes FingerMapper and ray-casting. For the temporal demand, hand tracking and ray-casting received a higher score than attach and direct. Figure 13b depicts the results of avatar embodiment (from -3 to 3). FingerMapper has a neutral score in the ownership subscale for attach, and a neutral toward slightly negative score for direct. For the agency subscale, the attach and direct mapping has a slightly positive score. For the location subscale, both FingerMapper received a neutral score. Figure 13c shows the scores of the presence questionnaire, where FingerMapper received scores around average (4 out of 7).}
  \label{fig:study-questionnaire}
\end{figure*}

Figure \mbox{\ref{fig:study-boundingbox}} shows the physical position of all participant's index fingertips from two perspectives (HMD movement included). We observed the pattern of reaching toward the target layout. Participants' movements were closer to the center in \mbox{\FingerMapper} functions and \mbox{\raycastingCond} than \mbox{\physicalCond}. Black points represent the HMD position. Since we conducted Study 1 in an open space without any constraints, participants still had some body movements. 
% bonding box includes index finger positions from all participants: want to see the maximum
We calculated these maximum bounding cubes to indicate interaction volume, showing how much space the participants used with each \mbox{\textsc{technique}}. These volumes allowed us to quantify how much we could reduce this volume. The \mbox{\physicalCond} had the largest interaction volume ($0.888\ m^3$) followed by \mbox{\Direct} ($0.782\ m^3$), \mbox{\Attach} ($0.602\ m^3$) and \mbox{\raycastingCond} ($0.506\ m^3$). \mbox{\Attach} could reduce the interaction volume by 32.2\% to be comparable to \mbox{\raycastingCond}.
% Calculate this data by participants and conditions to see if there is a difference between four conditions on individual interaction volume.
\hl{Also, we analyzed the interaction volume data by participants and conditions. By performing a logarithmic transformation, we found a statistically significant difference between \mbox{\textsc{technique}} ($F_{3, 36} = 32.8, p < .01, \eta^2 = 0.73$). In a post-hoc analysis, \mbox{\physicalCond} ($M = 0.44, SD = 0.14$) had a statistically significant higher volume than \mbox{\Attach} ($M = 0.21, SD = 0.1$), \mbox{\Direct} ($M = 0.22$, $SD = 0.12$), and \mbox{\raycastingCond} ($M = 0.11, SD = 0.05$) ($p < .01$ for three comparisons). There was no difference between \mbox{\Attach} and \mbox{\Direct}, but both had a statistically significant higher volume compared to \mbox{\raycastingCond} (Ray-Attach: $p = .02$, Ray-Direct $p = .04$).}

\textbf{Fatigue and NASA-TLX.}
Figure \ref{fig:study-questionnaire}a shows the results of NASA-TLX. We were particularly interested in the mental and physical demand, and our analysis revealed significant differences (Mental: $\chi^2_{3} = 24.64, p < .01, \eta^2 = 0.63$, Physical: $\chi^2_{3} = 19.73, p < .01, \eta^2 = 0.51$). The mental demand of \mbox{\physicalCond} ($M = 1.62, SD = 1.19$) and \mbox{\raycastingCond} ($M = 2.31, SD = 1.80$) were significantly lower than \mbox{\Direct} ($M = 4.62, SD = 1.94$) and \mbox{\Attach} ($M = 4.54, SD = 2.15$) ($p < .05$). For the physical demand, \mbox{\physicalCond} ($M = 5.85, SD = 0.90$) had a higher score compared to \mbox{\Attach} ($M = 3.85, SD = 1.14$), \mbox{\Direct} ($M = 4.05, SD = 1.75$), and \mbox{\raycastingCond} ($M = 2.54, SD = 1.66$) because participants had to reach out their arm in the \mbox{\physicalCond} condition. Here, comparisons between Attach-Hand and Ray-Hand found significant differences in the post-hoc analysis ($p < .05$). For the other four subscales in NASA-TLX, there were significant differences in temporal demand, effort, and frustration (Temporal: $\chi^2_{3} = 13.27, p < .01, \eta^2 = 0.34$, Effort: $\chi^2_{3} = 12.20, p < .01, \eta^2 = 0.31$,  Frustration: $\chi^2_{3} = 8.17, p = 0.043, \eta^2 = 0.21$), and there was no significant effect on the subscale of performance ($\chi^2_{3} = 6.36, p > .05, \eta^2 = 0.16$). One significant mean difference in temporal demand was found in the post-hoc comparison where \Attach\ ($M = 2.92, SD = 1.44$) has a lower score than \physicalCond\ ($M = 4.38, SD = 1.39$) ($p < .05$). 

\textbf{VBO and Presence.}  \autoref{fig:study-questionnaire}b shows the score of three subscales for VBO. The values were averages of individual questions and ranged between -3 and 3. A score higher than zero represents the feeling of VBO, and a score lower shows the lack thereof. There were significant differences in \textsc{techniques} on three subscales we measured (Ownership: $\chi^2_{3} = 28.97, p < .01, \eta^2 = 0.74$, Agency: $\chi^2_{3} = 18.22, p < .01, \eta^2 = 0.47$, Location: $\chi^2_{3} = 14.56, p < .01, \eta^2 = 0.37$). For the post-hoc comparison in the ownership subscale, we only found \mbox{\physicalCond} ($M = 2.49, SD = 0.46$) was significantly higher than \mbox{\Attach} ($M = 0.51, SD = 1.14$) and \mbox{\Direct} ($M = -0.39, SD = 1.59$) ($p < .01$). For the agency subscale, \mbox{\physicalCond} ($M = 1.65, SD =0.58$) was significantly higher than \mbox{\Direct} ($M = 0.48, SD = 1.17$) and \mbox{\Attach} ($M = 0.83, SD = 0.66$) ($p < .05$). No significant mean difference was found in the post-hoc comparisons of the location subscale.

Finally, \mbox{\autoref{fig:study-questionnaire}c} shows the General Presence in IPQ. Please see the supplementary material for all subscales (Spatial Presence, Involvement, and Experienced Realism). There were significant differences in General Presence ($\chi^2_{3} = 14.51, p < .01, \eta^2 = 0.37$) and Experienced Realism ($\chi^2_{3} = 11.5, p < .01, \eta^2 = 0.30$).  No mean differences were found in the post-hoc comparison of General Presence and Experienced Realism. No significant difference was revealed in Spatial Presence ($\chi^2_{3} = 4.42, p = .22, \eta^2 = 0.11$) and Involvement ($\chi^2_{3} = 3.92, p = .27, \eta^2 = 0.10$).

% Finally, Figure \ref{fig:study-questionnaire}c shows the General Presence in IPQ. For all the subscales (Sptaial Presence, Involvement, and Experienced Realism), please see the supplementary material. There was a significant difference in the overall IPQ score ($\chi^2_{3} = 14.51, p < .01$), but no significant mean difference was found in post hoc comparisons.

\subsection{Discussion}
In the following, we position both \FingerMapper\ mapping functions between two extremes: {\physicalCond} and {\raycastingCond}.

\textbf{Performance.} 
We found using \FingerMapper\ leads to a significant increase in the task time. Parts of this can be explained by the learning curve that we observed. Despite the initial training phase, participants still showed improvements during the study. Additionally, {\Attach} and {\Direct} showed a higher mental demand, effort, frustration compared to {\physicalCond} and {\raycastingCond}. Participants also tended to a lower performance\footnote{The scale of performance is reversed in NASA-TLX. A low-performance score means the user felt perfect considering how successful (or satisfied) while performing a task, and a high-performance score means the user felt failure while doing a task.} using \FingerMapper. This is probably due to the novelty of the technique and the initial phase, where users needed to get used to the different mappings. Overall, we argue the task time is still within an acceptable range since eight participants reported in the questionnaire that they would use one of the finger mappings to interact in confined spaces (two \Direct, six \Attach, and five \raycastingCond\ over 13 participants). 

\textbf{Motion and Fatigue.}
Our goal was to reduce fatigue, physical motion, and interaction volume for virtual arm interaction in a confined space. Results indicated that \FingerMapper\ had significantly lower scores on the physical demand than {\physicalCond}. We also found that {\raycastingCond} had an even lower score, positioning \Attach\ and \Direct\ in between both of them.

\Attach\ reduced the necessary physical path length down to approximately 67.8\%, positioning it again between \textit{\raycastingCond} (57.0\%) and \textit{\physicalCond} (considered here as 100\%). Additionally, we found that \textit{\Attach} reduced the overall interaction volume compared to \textit{\physicalCond} by 32.2\%. \hl{For the average of the individual interaction volume, \mbox{\Attach} reduced this metric down to 46.7\% of \mbox{\physicalCond}.} 
% We argue \textit{\Attach} (0.602 $m^3$) has an almost similar interaction volume as \textit{\raycastingCond} (0.506 $m^3$) while allowing the case to perform large virtual arm motions.
We argue \Attach\ (0.602 $m^3$) and {\raycastingCond} (0.506 $m^3$) have similar maximum interaction volumes while allowing the case to perform large virtual arm motions in confined spaces. Also, {\raycastingCond} has substantially less DoF, requiring additional extensions to allow for rotation and retraction (e.g., controlling the depth of the ray \cite{laviola3DUserInterfaces2017}). Without these extensions, {\raycastingCond} cannot enable manipulation as \FingerMapper. While {\physicalCond} had little additional virtual motions, both \FingerMapper\ and {\raycastingCond} showed a similar (\autoref{fig:study-results-quantative}c) level of additional virtual distance traveled, indicating participants had a quick burst in the direction of the target with overshooting and had an additional correction phase afterward.

\enlargethispage{5pt}

\textbf{VBO and Presence.}
Here we mainly consider {\Attach}, {\Direct}, and {\physicalCond} because the concept of ``body ownership'' is, in general, difficult to apply to an abstract pointing device such as a ray \cite{linNeedHandHow2016}. However, to control the visual of virtual arms in each condition, we used a tracked visualization of the user's hand. This probably led some participants to rate VBO toward the hand and not the actual ray as we intended. We still report the VBO score of {\raycastingCond} for completeness, but we do not provide further interpretation of this score.

Initially, we expected {\Attach} and {\Direct} can achieve a VBO score similar to {\physicalCond}. However, {\Direct} could not preserve any degree of VBO compared to {\Attach}. Although {\Direct} still preserved some VBO in the agency subscale, it was dominated by the visual of virtual arms. Therefore, {\Direct} had neutral or even negative scores in the ownership and location subscales. It was less preferred when asked whether participants would use it in a confined space. Nevertheless, {\Attach} could preserve a general sense of ownership, agency, and even location scores. 

In General Presence, all conditions received an average score above four on the 7-point scale. We believe all \mbox{\textsc{techniques}} could keep a rather high level of presence and did not break the user experience because General Presence scores were all above four. Nevertheless, \mbox{{\physicalCond}} had the highest score followed by \mbox{{\raycastingCond}}, \mbox{{\Direct}} and \mbox{{\Attach}}. In Spatial Presence and Involvement, four techniques did not show statistically significant differences in their score. The reason might be all participants were in the same virtual environment and interacted with the same task. Experienced \nobreak Realism ranged between three to four, probably caused by the \nobreak virtual environment being almost clear for study control.

\textbf{Observations and Qualitative Feedback.}
We observed participants with little to no VR experiences tended to express frustration when they could not fully control \FingerMapper\ during practicing. They also preferred {\raycastingCond} for using VR in a confined space. Since the task in the study can be achieved in a simple way physically (i.e., hand tracking), spending extra effort to complete this easy task with \FingerMapper\ indeed troubled some participants. 

The participants with more VR experiences could learn and control the mapping faster than others. One participant who frequently uses VR for gaming mentioned (P13), \textit{``At the beginning, I struggled a little with controlling, but once I understood the mapping, I think I can use this mapped arm to do more VR interaction.''} \FingerMapper, despite a higher task time, has lower fatigue and physical motion. Results show that \textit{\Attach} can preserve the user's presence and partially some degree of VBO. 

%%%%%%%%%%%%%%%%%%%%%%%%%%%%%%%%%%%%%%%%%%%%%%%%%
\section{Study 2: User Experiences in the Confined Space}
% we would also need to tell what the goal of the study was
% what did we aim to find out
Study 1 was to understand the exact interaction time and experience metrics in a controlled environment. In Study 2, we wanted participants to experience using \FingerMapper\ inside a confined space, compared to \physicalCond, the current state-of-the-art for VR interaction. Our focus was on experience metrics (presence, enjoyment) and safety metrics (perceived safety, amount of collisions). 

\subsection{Apparatus and Setup}
\begin{figure}[t]
% \centering
\includegraphics[width=\columnwidth]{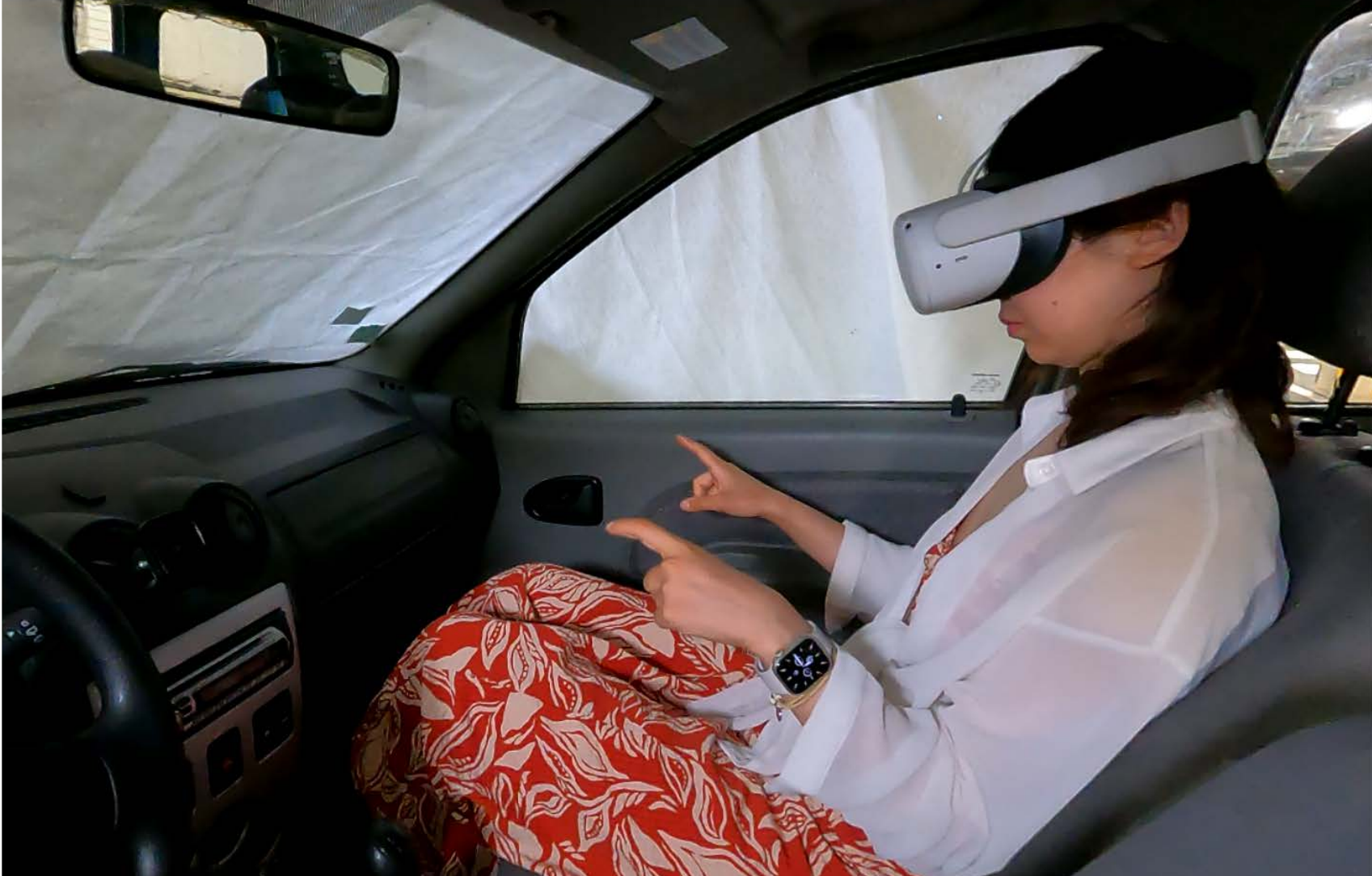}
 \caption{A participant seated in the front passenger seat.}
 \Description{Figure 14 shows the setup of Study 2, where participants sat inside the front passenger seat of a car and played FingerGrabber with Attach and hand tracking.}
 \label{fig:study2-setup}
\end{figure}

While the setup in study 1 was in an open space, study 2 was in the front passenger seat of a standing car (Figure \ref{fig:study2-setup}, Dacia Logan MCV) parked in our institution. The main goal was to test our technique inside a realistic representation of a confined space. We used \mbox{\FingerGrabber} deployed on an Oculus Quest 2 for the scenario because it involves object interaction. For the techniques, we used the same version of \mbox{\physicalCond} and \mbox{\Attach} in Study 1. We chose \mbox{\Attach} because it outperformed \mbox{\Direct} in VBO and performance. Participants were seated on the front passenger seat, using the stationary boundary of the Oculus Quest. The distances from the chest to the surroundings were roof: 50 cm, windshield: 85 cm, right car door: 40 cm, and left car door: 110 cm. 

\subsection{Study Design}
The study was a within-subject design with one independent variable \textsc{techniques} having two levels (\Attach\ and \physicalCond). Each participant had to experience \FingerGrabber\ (\autoref{sec:FG}) with {\Attach} and {\physicalCond}. \FingerGrabber\ aims to simulate VR experiences similar to Job Simulator, where the virtual objects would scatter within the arm space of the user for selection and manipulation tasks. These VR applications may be unavailable in a confined space because the virtual objects may be out of reach. Here, we argue that {\Attach} could be a good solution, reducing physical movement to keep the user safe during the interaction and preserving presence and enjoyment.

% explaining why not ray casting or other pointing techniques.
We did not include pointing techniques, such as ray-casting, since they require adapting the game or application. We envision \FingerMapper\ as an alternative to generate input stimuli for virtual arm interactions that can be used across all current applications relying on full-scale hand tracking without modifying applications.  

The order of the two conditions was counterbalanced. To measure the user experience, the participant had to fill out a questionnaire after each condition, including SUS presence questionnaire \cite{usohUsingPresenceQuestionnaires2000}, avatar embodiment \cite{gonzalez-francoAvatarEmbodimentStandardized2018} for VBO, and Intrinsic Motivation Inventory (IMI) \cite{mcauleyPsychometricPropertiesIntrinsic1989} for enjoyment. We added two 7-point Likert scale questions for perceived safety in the confined space: ``\textit{I feel safe when I was in the game,}'' and ``\textit{I feel worried about touching the car interior or other passenger when playing.}'' The study was video recorded. Therefore, we observed and counted collisions in each video footage to see if {\Attach} could reduce physical space for VR interaction. A collision was defined as the participant physically contacting the interior parts of the car. At the end of the study, participants had to rate (7-point Likert scale) which interaction techniques they prefer for using VR confined spaces.

\subsection{Procedure}
Participants were introduced to the study and signed the consent form. All participants were new to the system and did not attend Study 1. To reduce the ``wow'' effect, each participant practiced with the system in a training session prior to the data collection. They were free to practice \mbox{\Attach} or \mbox{\physicalCond} in an open space until they felt comfortable with the interaction (on average five minutes) until they could successfully select and put the target at the right place three times in a row. This step ensured each participant had a similar level of understanding and skill despite their prior VR experience. When they sat in the car, we gave the following instruction, ``\textit{imagine you are sitting inside a car of your friend (or a family member), please adjust the seat into a comfortable position as you usually do.}'' After adjusting the seat, the participant put on the Oculus Quest 2. The game lasted two minutes. Here, the goal was not to run performance-oriented evaluation (done in Study 1) but to observe users interacting in a realistic environment. The impact of the confined space (collisions) and perceived safety were independent of the duration of exposure. Collisions will happen right away, affecting the perceived safety. After each condition, they filled out the prior mentioned questionnaires in VR. When the participants completed both \textsc{techniques}, the experimenter interviewed them on their overall experience in VR and their preference of \textsc{techniques}. The whole study took an hour to complete.

\subsection{Participants}
Participants (n=13, six females and seven males) ages ranged from 18 to 33 ($M = 26.0, SD = 4.0$). All participants were recruited from our institute using convenience sampling. On a 5-point Likert scale, participants rated their expertise in VR between below average and average ($M = 2.2, SD = 0.8$). The frequency of using VR was rated between almost never and sometimes ($M = 1.8, SD = 0.8$). Overall, the sampled participants were mostly novice VR users. Participants reported normal or corrected vision and no motor impairments. The study was carried out following local health and safety protocols. Each participant was compensated with 10\texteuro\ for their participation.

\subsection{Results}
\begin{figure*}[t]
  \centering
  \includegraphics[width=\textwidth]{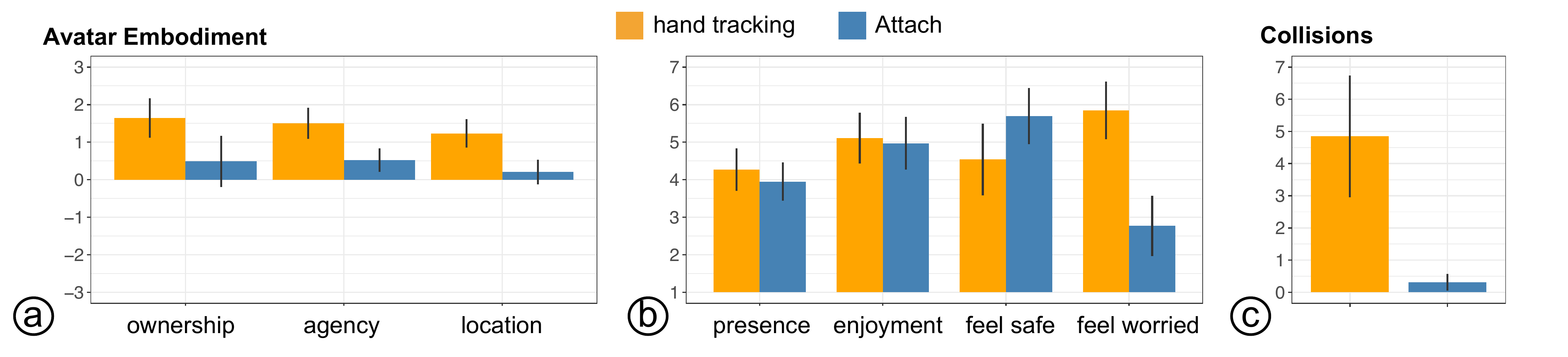}
  \caption{The results of: (a) virtual body ownership; (b) presence, enjoyment, and perceived safety; and (c) collision counts (Hand Tracking: M=4.9; \FingerMapper: M=0.3). The errorbar represents the 95\% CI.}
  \Description{Figure 15a shows the results of avatar embodiment on three subscales: ownership, agency, and location. Hand tracking received a higher score than Attach. However, Attach still preserves a small embodiment. Figure 15b depicts the results of a 7-point Likert scale on the presence, enjoyment, and perceived safety. Attach has a similar presence and enjoyment compared to hand tracking. Besides, Participants felt safer and less worried when using Attach compared to hand tracking. Figure 15c indicates that Attach has significantly fewer collisions compared to hand tracking in Study 2.}
  \label{fig:study2-result}
\end{figure*}

\autoref{fig:study2-result}a shows {\Attach} received a lower score in VBO compared to {\physicalCond}. Dependent t-test found \mbox{\physicalCond} had a higher score in ownership ($Mean_{Diff} = -1.15$, $95\%\ CI[-1.92, -0.38]$), agency ($Mean_{Diff} = -0.98$, $95\%\ CI[-1.55, -0.41]$, and location ($Mean_{Diff} = -1.03$, $95\%\ CI[-1.53, -0.52]$) of the avatar embodiment questionnaire. In each subscale, we revealed a significant difference (Ownership: $t_{12} = -3.27, p = 0.007, Cohen's\ d=0.90$; Agency: $t_{12} = -3.73, p = 0.003, Cohen's\ d=1.03$; Location: $t_{12} = -4.44, p = 0.0008, Cohen's\ d=1.23$). However, the VBO scores of {\Attach} were still above neutral, which is in line with the results in Study 1. Keep in mind that the scale of the avatar embodiment questionnaire ranges from -3 to 3. Our results would be equivalent to a rating of above 4 (above average) on a 7-point Likert scale. 

For presence (\autoref{fig:study2-result}b), a paired t-test found no statistically significant difference between \textsc{techniques} ($Mean_{Diff} = -0.3$, $95\%\ CI\ [-0.9, 0.3]$, $t_{12} = -1.21, p = 0.25, Cohen's\ d=0.34$). Although there was also no statistical significance in enjoyment ($Mean_{Diff} = -0.1$, $95\%\ CI\ [-0.7, 0.5]$, $t_{12} = -0.51, p = 0.62$, $Cohen's\ d=0.14$), both {\Attach} ($M=5.0$) and {\physicalCond} ($M=5.1$) received positive scores, showing participants enjoyed using \FingerGrabber. 

Our results show that {\Attach} benefited from having less physical movement that resulted in fewer collisions with the environment compared to {\physicalCond}. Re-watching all the video recordings of the participants, one author coded each instance (location, time) of a collision with any part of the car. We found that \textit{\Attach} had fewer collisions ($M=0.3, SD=0.5$) compared to {\physicalCond} ($M=4.9, SD=3.5$) (\autoref{fig:study2-result}c). A dependent t-test found a significant difference between the two techniques ($Mean_{Diff} = -4.5$, $95\%\ CI\ [-6.5, -2.5]$, $t_{12} = -5.02, p = .0003$, $Cohen's\ d=1.39$). Most of the collisions happened on the right car door (84\%). Additionally, we found a difference in the participant's perceived safety scores between the two \textsc{techniques}. \autoref{fig:study2-result}b shows the response to ``\textit{I feel safe when I was in the game,}'' and ``\textit{I feel worried about touching the car interior or other passenger when playing.}'' For ``feel safe'', participants had a higher rating in \Attach\ (M=5.7) than \textit{Hand} (M=4.5). For ``feel worried'', \Attach\ (M=2.8) received a lower score than \textit{Hand} (M=5.8). There were significant differences of the feeling of safety ($Mean_{Diff} = 1.2$, $95\%\ CI\ [0.3, 2.0]. t_{12} = 2.96, p = 0.01, Cohen's\ d=0.82$) and feeling of worry ($Mean_{Diff} = -3.1$, $95\%\ CI\ [-4.0, -2.2]. t_{12} = -7.41$, $p < .001, Cohen's\ d=2.05$). These results show that participants perceived a higher safety while using \Attach. Finally, 10 out of 13 preferred to use \Attach\ (\FingerMapper) over {\physicalCond} when asked which technique they would like to use for VR applications in confined spaces.

\subsection{Discussion}
In the confined space, our results show that \FingerMapper\ (\Attach) benefited from its fewer physical movements and kept presence and enjoyment comparable to \physicalCond. Most participants (10 out of 13) mentioned that reducing collisions was their main concern, and they preferred using \FingerMapper\ because it requires less physical space, so they feel safer. \FingerMapper\ was able to almost completely avoid collisions. In our study, only four participants collided one time with the right door at the beginning of the condition. P5 mentioned, ``\textit{FingerMapper respects the physical environment around you. So you make small movements and do not have collisions with the physical space. This way, we can be really into the VR experience but [avoid] the experience of colliding with something that's bringing you back to physical reality.}'' The subjective rating on perceived safety and preferences also reflected that participants preferred \FingerMapper\ as the interaction for confined spaces. 

Participants were aware of crossing over the space of the driver's seat. They felt that hand tracking is almost impossible to be applied in a confined space. P9 said, ``\textit{if I touched the door or window, I know it's not that dangerous because if I touch them, there won't be any accidents. However, if I know there's a driver on the left-hand side, Then I will avoid [going] too far with my arms.}'' P10 also mentioned, ``\textit{If a ball refreshed at the same position where I know, I probably will touch the interior of the car to grab the ball. If some people are sitting there and I touched those people, I probably will just avoid grabbing the ball in that direction and try to do the other direction.}'' Participants mentioned that during the interaction they had a mental model of the physical environment they were in, and rather avoided an interaction than collide with the environment. 

\FingerMapper\ can overcome the confined space and presents a safe (perceived safety, amount of collisions), ergonomic (less physical motion in Study 1), and enjoyable interaction (enjoyment) while preserving presence and VBO. Overall, participants' subjective rating showed a strong preference towards using \FingerMapper\ in confined spaces for VR interaction ($M=5.3, SD=0.9$) compared to \physicalCond\ ($M=3.2, SD=1.5$).

%%%%%%%%%%%%%%%%%%%%%%%%%%%%%%%%%%%%%%%%%%%%%%%%%%%%%%%%%%%%%%%%%%%%%%%%%%
\section{Lessons Learned}
As VR technologies become more mobile, users are able to engage in immersive virtual environments wherever they want. The confined space (e.g., commuting in a passenger seat) would be one of the new contexts where it does not allow for whole-body motion. We can observe smartphone behavior to understand what these confined spaces could look like. People use a smartphone and smartphone games during commutes or in small spaces (e.g., a toilette booth). If VR becomes further distributed, users would also want to engage with immersive applications in such spaces. 

Despite having a higher task time and smaller VBO, Study 1 shows that \FingerMapper\ could reduce physical movements and fatigue compared to \physicalCond. In the confined space of Study 2, \FingerMapper\ had fewer collisions than \physicalCond\ and could enable a similar level of presence and enjoyment inside a VR experience. Additionally, participants had a significantly higher perceived safety using \FingerMapper. Although it took a short training phase to master, participants preferred using \FingerMapper\ for VR interaction in confined spaces. In the following, we discuss insights into \FingerMapper\ gained from the design process and two studies.

% Using the concept of beyond-real interaction in VR \cite{abtahi_beyond_2022} opens new possibilities to break out of our known body shapes \cite{leigh_morphological_2017, dewez_you_2022, rosa_supernumerary_2019}. In our iteration of \FingerClimber\, we explored a supernumerary arm version of \FingerClimber\ where the user has four virtual arms mapped to either the left or the right index and middle finger. Here, the user has more choices of grips to reach, but the arms also potentially increase the mental workload for coordinating the two additional limbs. Overall, \FingerClimber\ shows our technique transfers these large grabbing gestures to small finger motions. 

\paragraph{\textbf{Reducing Fatigue in Long Duration VR Experiences with FingerMapper}}
\FingerMapper\ does not necessarily have to be used inside a confined space. Current VR interactions start to optimize for ergonomics to enable a longer and more comfortable experience \cite{zielaskoEitherGiveMe2020a, montanomurilloErgOErgonomicOptimization2017, wentzelImprovingVirtualReality2020, evangelistabeloXRgonomicsFacilitatingCreation2021b}. Our approach can be an option when the user is tired and wants a lighter experience in terms of physical fatigue. This approach challenges the dominant notion that VR user needs to perform the same actions as they do in real life (e.g., walking long distances). \FingerMapper\ provides users the feeling as if they would perform these interactions without actually performing them.

\paragraph{\textbf{Context Adaptive Interaction: Confined Space and New Metrics}}
Through the exploration of \FingerMapper, we found that participants started to value perceived safety and avoid collisions while using VR in a confined space. Current VR interaction techniques are frequently optimized around metrics such as presence or performance. \FingerMapper\ demonstrated that participants preferred the interaction technique with lower performance metrics but higher safety. We argue that future research metrics of VR interaction techniques will start to include constructs such as safety, social acceptance, and public perception. 

%%%%%%%%%%%%%%%%%%%%%%%%%%%%%%%%%%%%%%%%%%%%%%%%%%%%%%%%%%%%%%%%%%%%%%%%%%
\section{Limitations and Future Works}
Although \mbox{\FingerMapper} can enable virtual arm interaction for VR in a confined space while having less physical movement and fatigue, fewer collisions, and higher perceived safety, there are still some limitations to be improved. First, the displacement between the fingers and virtual arms resulted in lower VBO. One potential solution could be adding haptic feedback at the fingertip (e.g., tactile feedback) when the virtual arm intersects with a target to enhance the overall embodiment. Second, we introduce \mbox{\Attach} and \mbox{\Direct}, but the mapping function can be designed in many different forms, giving rise to future work exploring different mappings.

The current studies focused on the controlled study and did not evaluate \FingerMapper\ inside a public confined space (e.g., a bus). The reason is that we were initially interested in comparing the individual techniques in a controlled environment to have a fair comparison to regular hand tracking. Future work could conduct further studies in public spaces to understand the social implications of finger mappings. \hl{Our studies explored the effect of FingerMapper by observing path length with a small sample size, which is one limitation of the current work. Future research could aim for a large sample size experimental setup to verify the effect of re-associating body parts in both position and rotational movement.}

To enable VR interaction in confined spaces, one can either adapt the application to require less physical movement \mbox{\cite{im-henriFingerSaber2022, melnickMiniatureBeatSaber2020}} or the input. \mbox{\FingerMapper} focused on adapting the input since it allows for having one implementation (might even be on the system level) that creates a mapping from a confined space interaction toward a whole-body interaction in VR. This allows the designers of VR applications to abstract from the context in which the application would be used and always design for a whole-body experience. However, the challenge of adapting the input is to find an appropriate mapping that has the same DoF while using less physical space and still being usable. This inevitably leads to trade-off decisions and creates a set of limitations for the proposed input adaptation. 

While FingerMapper provides a mapping for the basic input mechanisms of whole-body hand-tracking interaction (object selection and manipulation), our trade-off is losing efficiency (higher selection time, less precision). Additionally, the kinematic structure of fingers is different to the structure of arms (e.g., a throwing motion with arms is possible because of the DoF of the connecting joints, while the connecting joints of the finger have a different DoF). This results in some actions (e.g., throwing) being more difficult to perform using FingerMapper. However, we argue that FingerMapper focuses on enabling the basic functions of interacting in VR (object selection and manipulation) and presenting a mapping that can be generally applied to most applications and allow for a safe interaction in confined spaces. Future work should explore how complex actions can be mapped to small-space interaction techniques.

\section{Conclusion}
\FingerMapper\ enables whole-body VR interaction in VR, allowing users to control arm and hand interactions in confined spaces through small finger motions. Three applications highlight the different functionalities and scenarios, including large slicing motions, object interaction, and locomotion. Two studies show that \FingerMapper\ can reduce physical movements, fatigue, and collisions. Osur techniques reach a similar level of presence and enjoyment in the confined space with higher perceived safety and user preferences. We see \FingerMapper\ as an alternative to \physicalCond---if the user interacts in a confined space, and \raycastingCond---if the user needs more functionality and embodiment of the avatar. Overall, we argue input techniques in VR should be similarly easy to adapt to the context as the tracking options are (e.g., room-scale and stationary tracking). The concept of \FingerMapper\ would work with any application when the environment is not allowed for whole-body motion while preserving the VR user's experience and safety.

\begin{acks}
We thank our anonymous reviewers for their suggestions on improving this paper and all participants. This work was partially conducted within the Investments for the Future Program (PIA), CONTINUUM (ANR-21-ESRE-0030), funded by French National Research Agency (ANR) and the Living Lab 5G project financed by the Banque Publique d'Investissement (BPI).
\end{acks}

\balance
\bibliographystyle{ACM-Reference-Format}
\bibliography{FingerMapper}

% \appendix









\end{document}